\documentclass[a4paper,fleqn,usenatbib]{mnras}
\usepackage[T1]{fontenc}
\usepackage{ae,aecompl}
\DeclareMathAlphabet\mathbfcal{OMS}{cmsy}{b}{n}
\pdfoutput=1

\usepackage{graphicx}	% Including figure files
\usepackage{amsmath}	% Advanced maths commands
\usepackage{amssymb}	% Extra maths symbols

\usepackage{cuted}
\usepackage{lipsum}
\usepackage{bm}
\usepackage{lmodern}
\usepackage{pdflscape}
\usepackage{rotating}
\usepackage[bottom]{footmisc}
\usepackage{adjustbox}
\usepackage{multicol}
\usepackage{xcolor}
\usepackage{ulem}
\usepackage{soul}

\usepackage{capt-of}
\definecolor{dorange}{rgb}{1,0.4,0}

\title{AlgoSCR: An algorithm for Solar Contamination Removal from radio interferometric data}

\author[Phan et al.]{
Anh Phan$^{1}$\thanks{\href{mailto:anh@wisc.edu}{anh@wisc.edu}},
Santanu Das$^{1,2}$\thanks{\href{mailto:sanjone@gmail.com}{sanjone@gmail.com}},  
Albert Stebbins$^2$, 
Peter Timbie$^{1}$\thanks{\href{mailto:pttimbie@wisc.edu}{pttimbie@wisc.edu}}, 
Reza Ansari$^3$,
\newauthor
Shifan Zuo$^4$, 
Jixia Li$^5$, Trevor Oxholm$^1$, Fengquan Wu$^5$, 
\newauthor
Xuelei Chen$^{5,6,7}$, 
Shijie Sun$^{5,6}$, Yougang Wang$^5$, 
Jiao Zhang$^8$\\
\\
$^{1}$ Department of Physics, University of Wisconsin Madison, 1150 University Ave, Madison WI 53703, USA \\
$^{2}$ Fermi National Accelerator Laboratory, P.O. Box 500, Batavia IL 60510-5011, USA\\
$^{3}$ Universit\'e Paris-Saclay, CNRS/IN2P3, IJCLab, 91405 Orsay, France\\
$^{4}$ Department of Astronomy and Tsinghua Center for Astrophysics, Tsinghua University, Beijing 100084, P.R.China\\
$^{5}$ National Astronomical Observatory, Chinese Academy of Science,20A Datun Road, Beijing 100101, P. R. China\\
$^{6}$ University of Chinese Academy of Sciences Beijing 100049, P. R. China\\
$^{7}$ Center of High Energy Physics, Peking University, Beijing 100871, P. R. China \\
$^{8}$ School of Physics and Electronics Engineering, Shanxi University, Taiyuan 030006, P. R. China
}

\begin{document}

\maketitle

\begin{abstract}
Hydrogen intensity mapping is a new field in astronomy that promises to make three-dimensional maps of the matter distribution of the Universe using the redshifted $21\,\textrm{cm}$ line of neutral hydrogen gas (HI). Several ongoing and upcoming radio interferometers, such as Tianlai, CHIME, HERA, HIRAX, etc. are using this technique.  These instruments are designed to map large swaths of the sky by drift scanning over periods of many months. One of the challenges of the observations is that the daytime data is contaminated by strong radio signals from the Sun.  In the case of Tianlai, this results in almost half of the measured data being unusable. %Other radio %interferometers 
%telescopes around the world also face similar issues.
We try to address this issue by developing an algorithm for solar contamination removal (AlgoSCR) from the radio data. %Single dish instrument also suffer from solar contamination. However, in this paper we only 
The algorithm is based on an eigenvalue analysis of the visibility matrix, and hence is applicable only to interferometers. 
% focus on 
%because AlgoSCR is 
%, which is a collection of measurements of the interference pattern produced by different sets of two apertures. 
We apply AlgoSCR to simulated visibilities, as well as real daytime data from the Tianlai dish array.  The algorithm can remove most of the solar contamination without seriously affecting other sky signals and thus makes the data usable for certain applications.
\end{abstract}


\section{Introduction}

Cosmologists study the Universe on the largest observable distance scales in order to understand its origin and evolution. In the past few decades, cosmic microwave background (CMB) instruments have mapped almost the entire sky with high sensitivity and fine angular resolution. These maps measure the intensity and polarization fluctuations at the last scattering surface and remain a primary tool for studying the Universe.  %Experiments like WMAP, Planck have measured the CMB with exquisite precision. 
%{\color{red}VLA, MWA, CHIME, LWA, HERA daytime Sun removal. How big of a problem is the Sun signal} {\color{blue} I don't get any paper. But I need to check. Also we need to modify the introduction a lot+-. I don't see any papers}
However, for understanding the nature of dark matter and dark energy, it is essential to study the evolution of structure as a function of time. 
%CMB provides a two-dimensional map of the Universe during the last scattering surface.
 Galaxy redshift surveys have been extremely successful in mapping the large scale structure of the Universe by cataloging the distribution of luminous galaxies in redshift space. These maps can be used, for example,  to observe the characteristic baryon-acoustic oscillation (BAO) signal, which can be used as a standard ruler to extract cosmological parameters. However, as we map larger and more distant volumes of the Universe, the method faces multiple challenges.  For example,  the galaxies become fainter and spectral lines are redshifted to wavelengths that are difficult to detect from the ground.

Intensity mapping, a radically different technique, creates 3D maps using the $21\,\textrm{cm}$ emission of neutral hydrogen (HI) without resolving individual galaxies.  This line is unique in cosmology as, for $\lambda > 21\,\textrm{cm}$, it is the dominant astronomical line emission for all redshifts. Hence, to a good approximation the wavelength of a spectral feature can be converted to a redshift without having to first identify the atomic transitions. 
%A new technique called  (measurements of the 21 cm emission line from neutral hydrogen) can in principle be used 
In principle, HI intensity-mapping could be used to make 3D maps of matter at all redshifts up into the “dark ages” $(z \approx 100)$, even before galaxies have formed. 

The first HI intensity mapping observations began over a decade ago~\citep{Abdalla2005,Peterson2006, Morales2008, Chang2008,Mao2008} and interest has continued to grow \citep{ansari2018inflation, liu2020data, slosar2019packed}. A number of dedicated projects have been launched to detect the signal and turn the technique into a useful cosmological tool.  These are mainly interferometers such as CHIME~\citep{bandura2014canadian, newburgh2014calibrating}, Tianlai~\citep{chen2011radio,xu2014forecasts, das2018progress,li2020tianlai,wu2020tianlai}, MWA \citep{tingay2013murchison}, LWA \cite{eastwood2018radio}, HERA \citep{deboer2017hydrogen}, HIRAX~\citep{Newburgh:2016mwi}, and PUMA \citep{slosar2019packed}, but also include single dishes with multiple feed antennas, such as BINGO \citep{battye2012bingo,dickinson2014bingo,wuensche2019bingo} and FAST\citep{Hu_2020}. Intensity mapping instruments can address questions at a variety of redshift ranges. At $z\sim 10$ they probe the Epoch of Reionization (EoR), star formation, and galaxy assembly, while at lower redshifts they trace large scale structure for studies of dark energy, etc. \citep{peterson2006hubble,bull2015late,battye2004neutral,abdalla2005probing, chang2008baryon, Mao2008, Morales2008}.  

So far, the HI signal has not been detected using intensity mapping by itself.   Intensity mapping observations have set upper limits on HI from the EoR \cite{ali2015paper} and, in the post-recombination epoch, have detected HI when cross-correlated with galaxy redshift surveys~\citep{masui2010near,Masui2013,anderson2018low}. A number of challenging systematic effects must be overcome to allow autocorrelation detections. The foremost of these is separating the HI signal from Galactic and extra-Galactic astronomical foregrounds, which are $\sim 4-5$ orders of magnitude brighter \citep{Liu&Shaw2019}.  The Sun represents an astronomical foreground that is even brighter and of a different character. 
%Of the many foreground removal challenges, solar contamination of the data is one of the most difficult.

\begin{figure*}
  \centering
  \includegraphics[width=0.54\textwidth,trim = 0 150 0 50, clip]{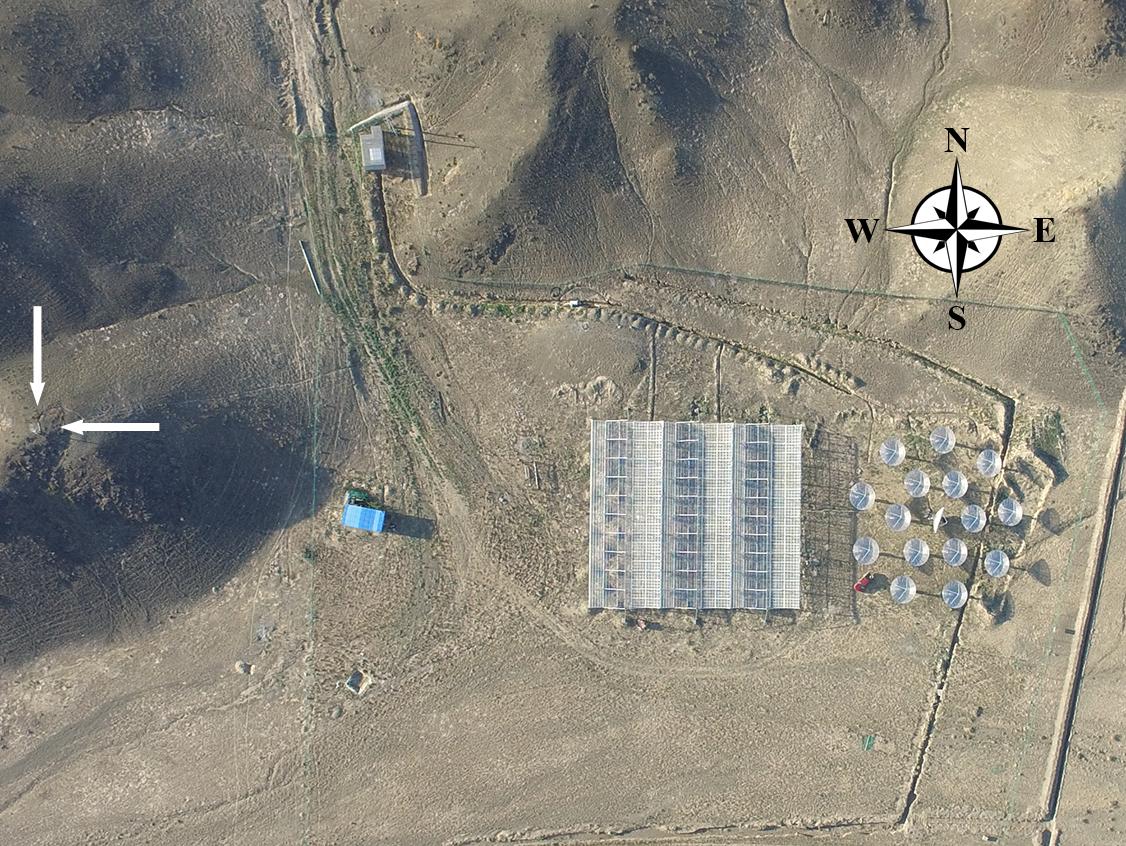}
  \includegraphics[width=0.43\textwidth]{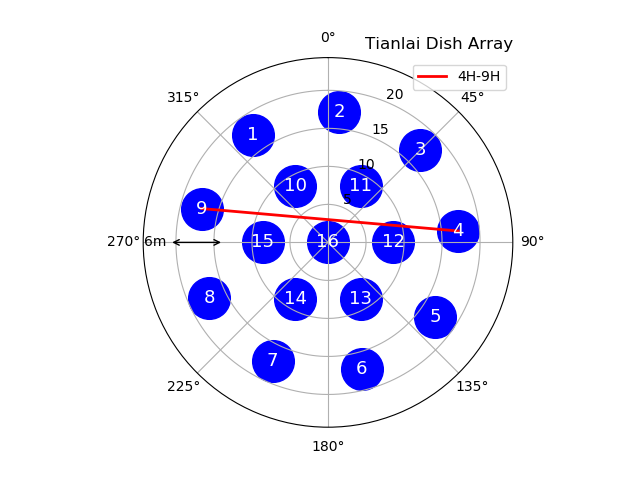}
  \caption{\textbf{Left:} A top view photograph of the Tianlai arrays, which consist of the Dish Array and the Cylinder Array. The photo was taken with a drone at a height of 280~m above the ground. The arrays saw first light in 2016. The position of the calibration noise source is indicated by the white arrows on the left. \textbf{Right:} 
  A schematic diagram of the Tianlai dish array. The dishes are arranged in two concentric circles of radius $8.8\,$m and $17.6\,$m around a central dish. The dishes have dual-linear polarization feed antennas with one axis oriented parallel to the altitude axis (horizontal, H, parallel to the ground)) and the other orthogonal to that axis (vertical, V). For example, red line shows one of the baselines that is the H polarization of dish 4 correlated with the H polarization of dish 9: [4H 9H]. The above image is reproduced from~\citep{wu2020tianlai}. %\rzorange{\st{This figure corresponds exactly to the dish paper figure 1. Either keep only the right panel (array configuration) or give the reference in the caption.}}
  }
  \label{fig:tianlai_array}
 \end{figure*}

The daytime data from radio interferometer arrays in general, and the Tianlai dish array in particular, are contaminated by the solar signal, making the data unusable for most astronomical analyses. The lost data have a significant impact on observing efficiency; reaching the required survey sensitivity means observing the sky for almost twice the number of days. This penalty is particularly problematic for HI intensity mapping, where long integration times (months or years) are necessary to detect the HI signal.  Furthermore, this data loss prevents obtaining continuous, 24~hr data sets, which allow  dense coverage of the $u-v$ plane and facilitate detection of periodic signals.
Furthermore, not having 24 hours of continuous usable observations prevents the application of m-mode map making techniques \cite{Shaw2013}.
The (u,v) coverage can still be quite good with nighttime data, and one can recover full 24 hours RA coverage by combining nighttime data from observations about 6 months apart. 
In this paper we try to remove the solar contamination from the daytime data from a radio interferometer. %As a part of the Tianlai collaboration, 
We have used the data from the Tianlai dish array as our test sample. However, the problem is not unique to Tianlai; the same algorithm may be used for other radio interferometric observations. While daytime observations with single dish radio telescopes are also plagued by the Sun's signal, this algorithm is only applicable to interferometer arrays.

The Tianlai Project is led by the National Astronomical Observatory of China (NAOC).  It consists of two pathfinder radio interferometers, an array of cylinder antennas and an array of dishes, at a radio-quiet site in Xinjiang, China \citep{chen2012tianlai,Li2020, wu2020tianlai}. The objective is to obtain high fidelity 3D images of the northern sky using HI intensity mapping. Construction of the Pathfinder was completed in 2016 and the arrays now undergoing commissioning. So far, we have collected over 258 days of data with the dish array and 114 days with the cylinder array and have begun the process of calibrating, making maps, and removing foregrounds. Both arrays are operated in drift-scan mode. Near-term analyses include cross-correlating the radio maps with galaxy redshift surveys. If successful, the arrays can be expanded to increase sensitivity. 

The analysis described in this paper concentrates on the dish array data. %The cylinder array is composed of three, fixed, cylindrical reflectors, each 40~m long and 15~m wide, equipped with 96 dual-polarization receivers. 
The Tianlai dish array consists of 16 steerable, 6~m diameter dishes; a schematic is shown in Fig.~\ref{fig:tianlai_array}, which also shows the dish numbering scheme. %is shown in the diagram. in the right. 
We use these dish numbers for referring to different baselines in the paper.  %For  each  array,  
The  feed  antennas,  amplifiers,  and  reflectors  are  designed  to  operate  from 600~MHz  to  1420~MHz. The  instrument  can  be  tuned  to  operate  in  an  RF  bandwidth  of 100~MHz  centered  at  any  frequency  in  this  range  by adjusting  the  local  oscillator  frequency in  the  receivers  and  replacing  the  bandpass  filters. The dish array currently operates between 685~MHz and 810~MHz, corresponding to redshift 0.75 < z < 1.07, divided into 512 equally spaced frequency bins of width 244~kHz. The 16 dual polarization feeds yield 32 autocorrelation visibilities and $32\times(32-1)/2=496$ cross-correlation visibilities, which are currently sampled every second. An additional electronic backend is being installed to search for transients (e.g. fast radio bursts) in parallel with the standard correlator used for HI mapping described here.  %to operate in parallel with the existing backend to sample faster (microseconds) to search for fast radio bursts.
The system noise temperatures for the dish antennas are $80 - 85$~K \citep{li2020tianlai,zhang2016sky,das2018progress,wu2020tianlai}.   

The objective of this paper is to describe an eigenvalue-based approach for removing solar contamination from radio interferometric data. We propose an algorithm, AlgoSCR, that can remove most of the solar contamination.  The paper is organized as follows. In the second section, we discuss the solar contamination problem in the Tianlai dish array in detail. The third and the fourth sections give the detailed algorithm for removing the solar contamination. We also show the results of our analysis on the real Tianlai data. To test what fraction of the Sun signal can be removed by our algorithm, and how much signal from other cosmic sources is removed by it, in section five we have applied it to simulated data where the amount of solar contamination and the external sources are known. %We show that the method can  remove the solar signal very effectively. 
In the discussion section we assess the efficacy of the method and the issues that we face when applying it. We also describe some future directions to pursue with this approach.

\section{The solar contamination problem}

%\rzorange{Would it be possible to add a cut through figure 3 beams, corresponding to the sun track. Include also a rough estimate of the sun brightness (around 750 Mhz) and expected Tianlai visibility signals in temperature. Compare then with figure 2. }

\begin{figure*}
    \centering
    \includegraphics[width=0.81\textwidth]{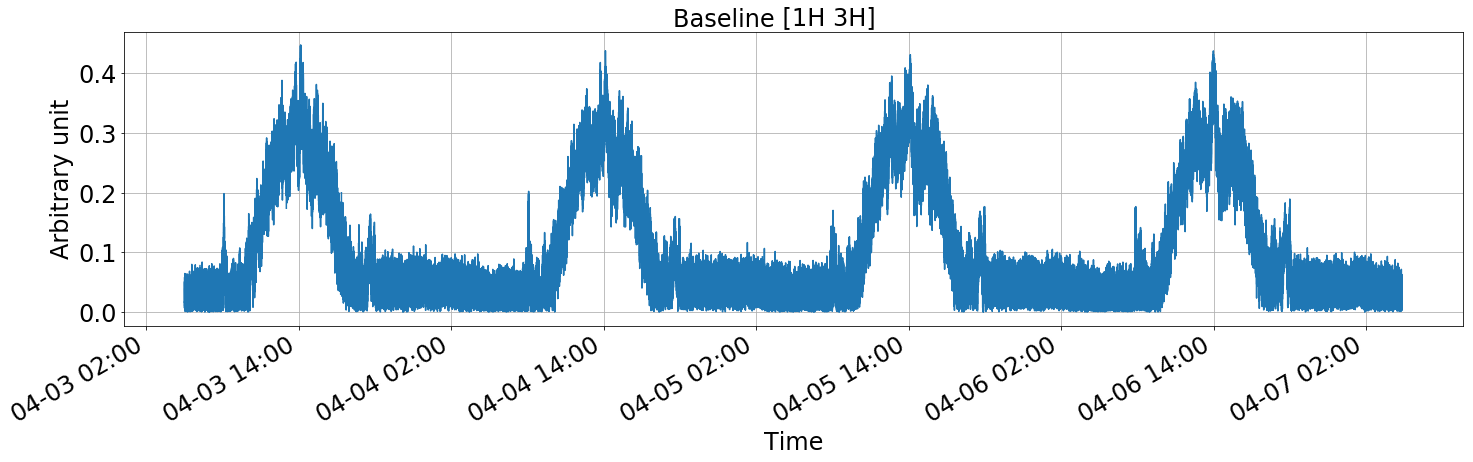}
    \includegraphics[width=0.81\textwidth]{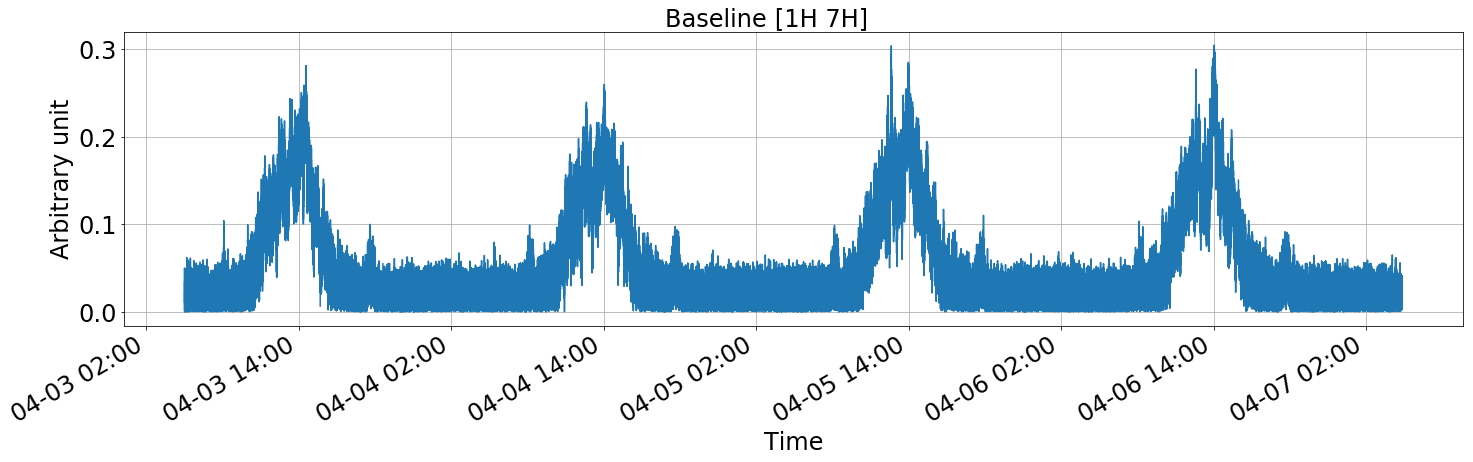}
    \includegraphics[width=0.81\textwidth]{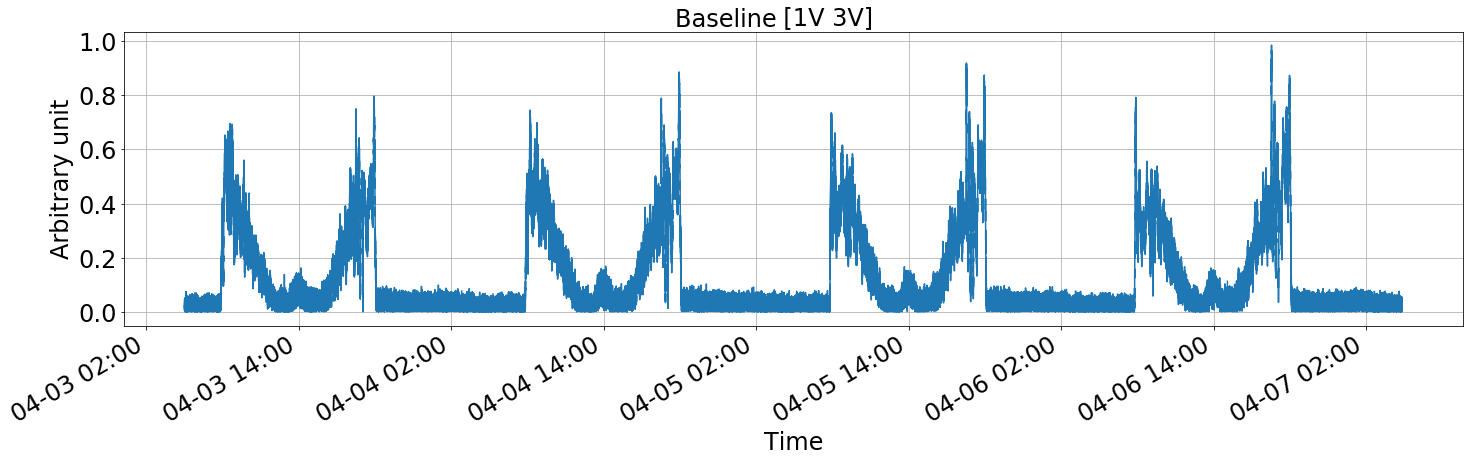}
    \includegraphics[width=0.81\textwidth]{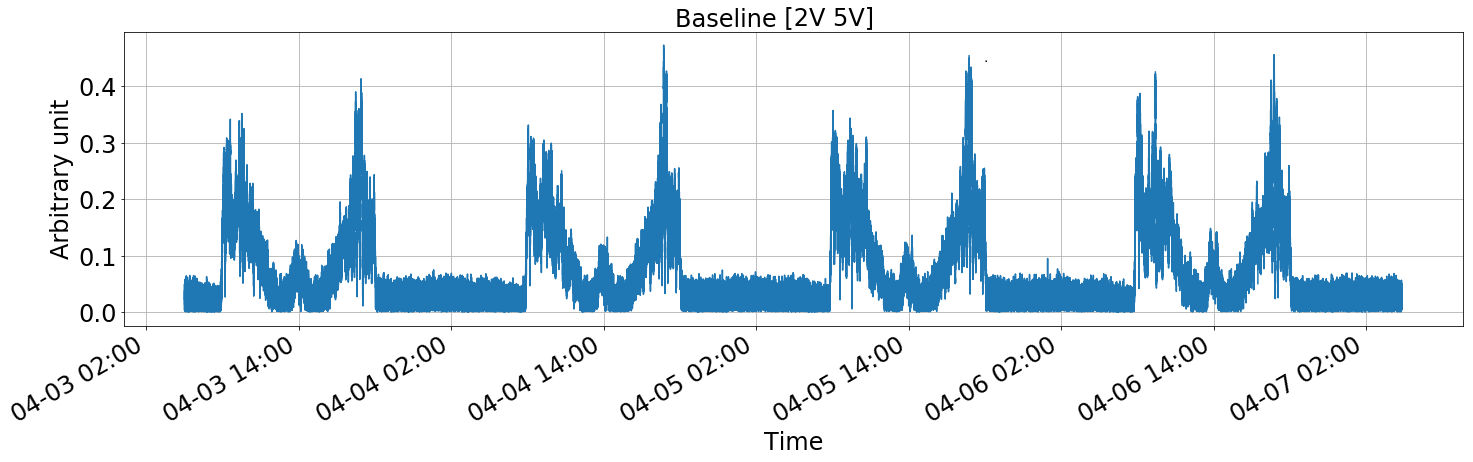}
    \caption{Value of the (uncalibrated) cross-correlation visibility amplitudes averaged over the 10 central frequency bands during 4 days of observations in April, 2019. Integration time is one second. %The visibilities are sampled once per second.
    Each plot corresponds to a different baseline, as indicated. The baseline numbering scheme appears in Fig.~\ref{fig:tianlai_array}. Time is given in local sidereal time.
    }
    \label{fig:Fig1}
\end{figure*}

%The issue with the solar contamination come into light when we analyze the north pole data from a 3 day continuous run of Tianlai telescope in 2017. 
The Tianlai data show strong contamination from the solar signal during the daytime. In Fig.~\ref{fig:Fig1} we plot the sum of the absolute visibility from 10 frequency channels (out of $512$ channels) at the center of the band, for 4 consecutive days (total 96 hours) for four different baselines.  The horizontal axis shows the time in hours, starting at the beginning of the observations. The 4 different plots are for 4 representative visibilities. We can see a roughly smooth visibility amplitude for about $10$ hours every day and then a sudden increase in the absolute visibility and a noisy pattern %in the visibility 
for about next $14$ hours. The smooth part of the visibility comes from the data taken during the night, whereas the noisy high amplitude section corresponds to daytime data. 

The plots clearly show that the daytime signal is several times stronger than the night. The shape of the contamination pattern also varies with baseline. Some of the baselines show a bumpy feature with the strongest visibility occurring near noon, whereas for other baselines the signal is strongest during Sunrise and Sunset and shows a `dip'  feature during the daytime. The top $2$ plots are the auto-polarization visibility corresponding to two horizontal feeds, whereas the bottom two plots show two auto-polarization visibility from two vertical feeds. The auto-polarization signals from similar feeds on other baselines show roughly similar types of patterns, except for a couple of baselines. The data are taken during observations of the North Celestial Pole (NCP) in April, 2019. During this observing period the path of the Sun is located at an angle of approximately  $85^\circ$ from the direction of the main beam. The plot gives an overview of the  magnitude of the solar contamination problem in the Tianlai dish array.  

\begin{figure*}
    \centering
    \includegraphics[width=0.45\textwidth,trim = 200 200 150 100, clip]{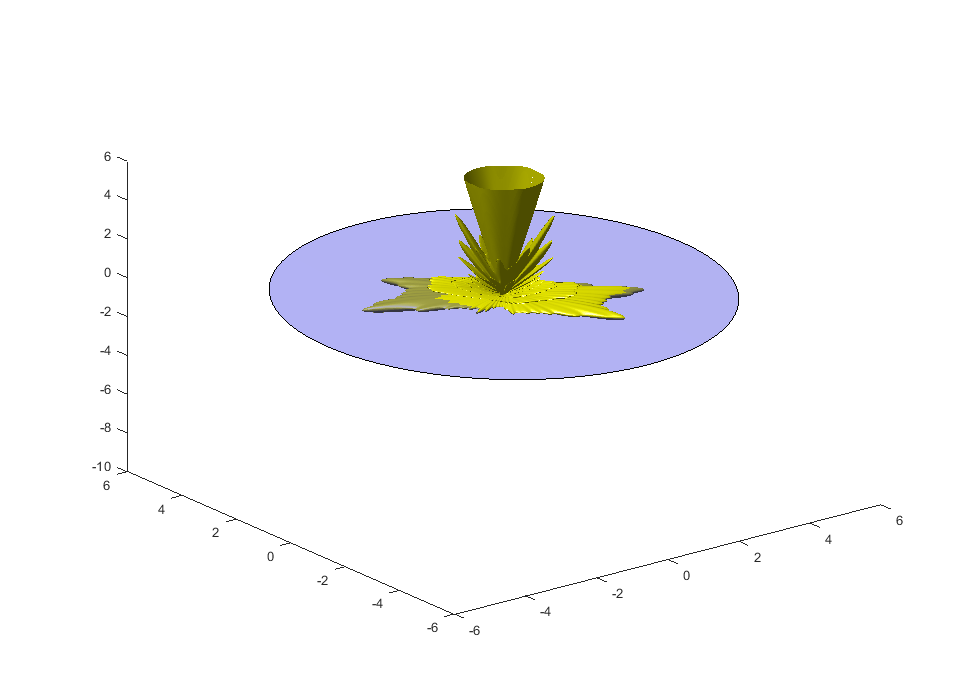}
    \includegraphics[width=0.45\textwidth,trim = 200 200 150 100, clip]{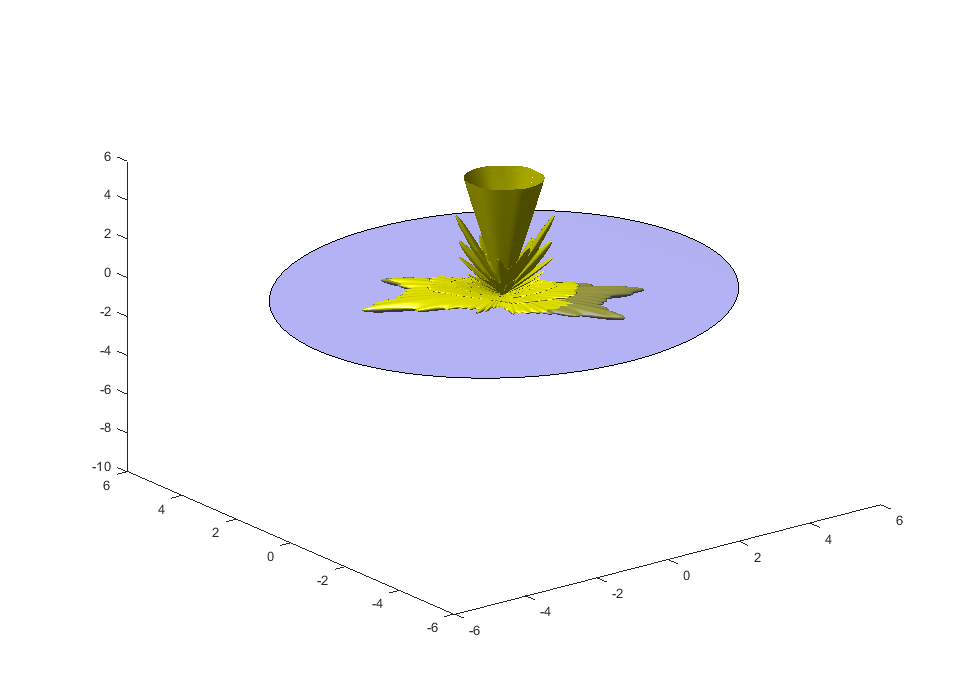}
    \caption{Plot of the beam pattern for a single dish in the Tianlai dish array, calculated using an electromagnetic simulation. The value of the antenna directivity is plotted in arbitrary linear units (not in dB). The main beam is oriented in the vertical direction and truncated %at 6
    in order to emphasize the sidelobes.  The path of the Sun is indicated by the blue plane for the observations analyzed here. The simulated beam shape is the same for the $V$ and $H$ polarizations, but one is oriented $90^\circ$ with respect to the other.  The plots show that the sidelobes are highly structured and the paths cross through  different sidelobes for the different polarizations.}
    \label{fig:fig2}
\end{figure*}
\begin{figure*}
    \centering
    \includegraphics[width=0.40\textwidth,trim = 0 0 0 0, clip]{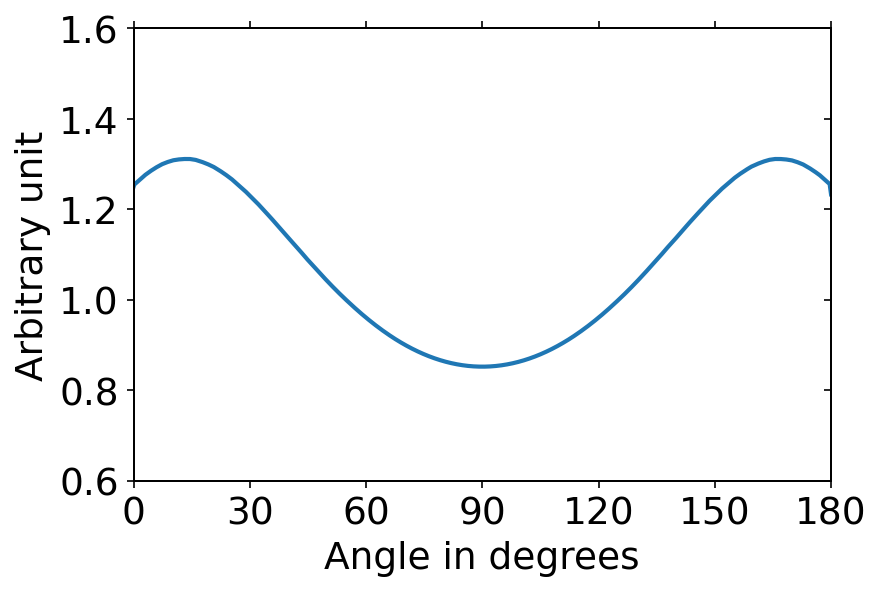}
    \includegraphics[width=0.40\textwidth,trim = 0 0 0 0, clip]{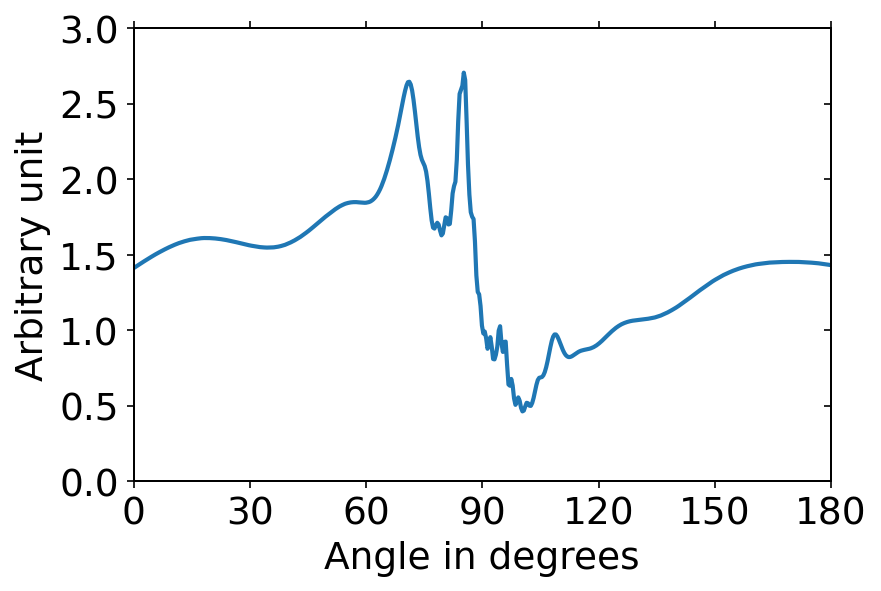}
    \caption{A cut through the Sun's track during daytime ($180^{\circ}$ corresponding to 12 hours) shown in  Fig.~\ref{fig:fig2}. The value of the antenna directivity is plotted in arbitrary linear units (not in dB). The plot shows that for one of the polarizations, the power is low during the daytime, whereas for the other polarization there is high power during certain parts of the daytime. Even though the plot on the right does not show the exact feature that we are getting during daytime (Fig. \ref{fig:Fig1}),  we must remember that the sidelobes from the electromagnetic simulations are not exactly the
    same as that of the real antenna.}
    \label{fig:cross-section}
\end{figure*}

\begin{figure*}
    \centering
    \includegraphics[width=0.47\textwidth]{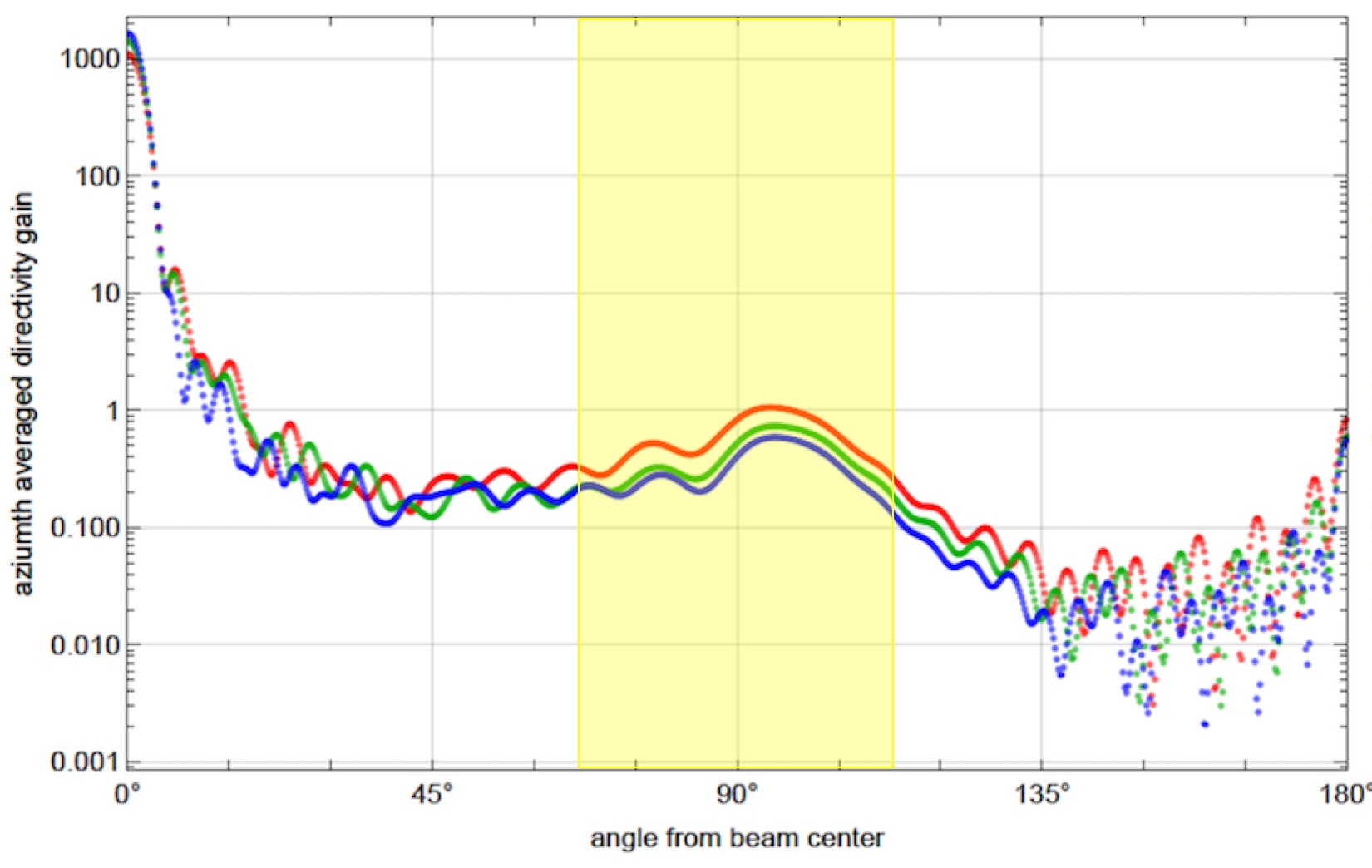}
    \caption{Simulated beam patterns as a function of beam angle $\theta$ from the beam center of the antennas for 3 different frequencies, 700 (red), 750 (green) and 800~MHz (blue). Each plot shows the absolute co-polar directive gain
    averaged over the azimuthal angle.  The angle is the polar angle calculated from the center of the beam. The yellow shaded region shows the range of polar angles for which the Sun appears in the sidelobes of the beam, ranging from $66.55^\circ$ at the Summer Solstice, to $113.45^\circ$ at the Winter Solstice. The gain is relatively flat over this range of angles and causes the Sun signal to vary by only a factor of about $6$ over the year.
    }
    \label{fig:beam_vs_theta}
\end{figure*}

%\begin{figure*}
%  \centering
%  \includegraphics[width=.47\textwidth]{figs/EastWestUAV%CSTsims.jpg}
%  \includegraphics[width=.47\textwidth]{figs/EastWestUAVCasAblowup.jpg}
%  \caption{Measurements and simulations of the H-plane antenna pattern of the V-polarization of one dish. Left: Pattern measured with the UAV compared with two electromagnetic simulations. Right: Patterns measured both by an UAV  and by a transit of Cas~A. In both figures the UAV flies in the E-W direction and the measurements and simulations are performed at 730~MHz.{\color{red} Do we need this plot?? I don't think its mentioned anywhere in the paper.}}
%  \label{fig:beam_vs_theta}
%\end{figure*}

An obvious conclusion of this strong daytime visibility is that the telescopes are responding to the Sun's illumination of their far sidelobes.  For the baselines measuring correlations of the $H$ polarization,  the antenna sidelobes are aligned with the direction of Sun near noon, providing a strong visibility at mid-day,  while for the $V$ polarization the Sun falls between two side lobes at noon, producing stronger signal during  Sunrise and the Sunset.  These effects are consistent with the expected responses of the feed antennas, which are essentially orthogonally oriented crossed dipoles. In Fig.~\ref{fig:fig2} we show part of the simulated beam pattern for a single dish measured using an electromagnetic simulation package (CST \footnote{\url{https://www.3ds.com/products-services/simulia/products/cst-studio-suite/}}).
We also show the plane of the path of the Sun using a blue circle for both polarizations under assumptions that the beam shape will be the same for both feeds. Even though the real shape of the beam function may be different from the simulated one, the plot can provide us an understanding about the origin of the particular type of shape seen by the telescope. The figure shows that for some polarizations the beam sensitivity in the Sun's path is highest during Sunrise and Sunset, and the path does not go through any other strong sidelobes throughout the day, whereas for the other beam, the path of Sun intersects some of the side lobes during midday, causing a strong response at noon.  In Fig.~\ref{fig:cross-section} we show  a cut through Fig.~\ref{fig:fig2} beams, corresponding to the sun track during daytime. We can see that for one of the polarization we are getting a low amplitude during the midday whereas for the other polarization (right plot) the amplitude is comparatively high during noon. %Even though we can see a sudden drop in power for few hours which is actually coming because that particular region falls between two side-lobes. 
The simulated patterns shown in Fig.~\ref{fig:cross-section} doesn't exactly replicate the observed pattern of Fig~\ref{fig:fig2}, because the sidelobes from these EM simulations don't exactly match that of the real beam. The sidelobes at this particular angle are also highly cluttered. A couple of degrees changes in the path gives rise to a very different shape in the sidelobes, making it difficult to reconstruct the exact pattern through such EM simulation.

\begin{figure*}
    \centering
    \includegraphics[width=0.8\textwidth]{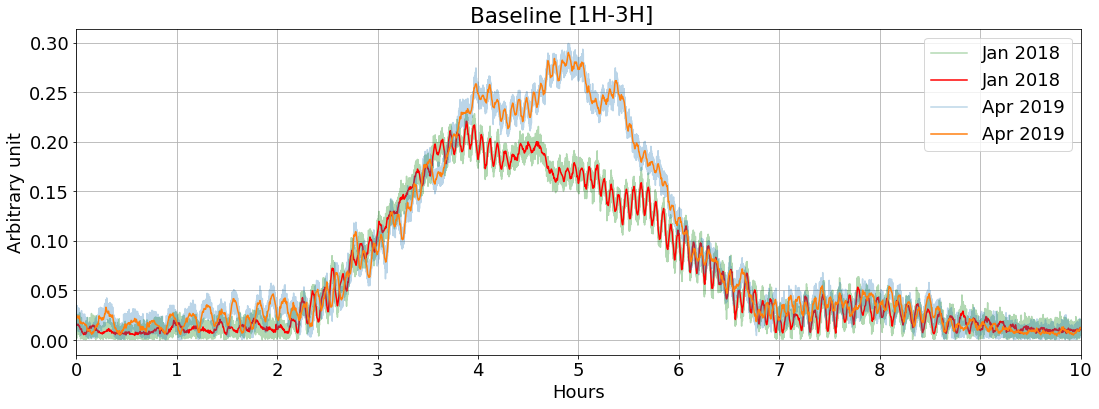}
    \caption{
    The amplitude of the daytime visibilities in January, 2018 and April, 2019. Due to the difference in the time of Sunrise in January and April, the 0 hr of each curve is adjusted so that the Sun signals from both data sets peak at about the same time. The fast oscillation fringes are seen in both observations. The green and blue curves are the amplitude of the visibility obtained from the telescope for January 2018 and April, 2019, respectively. The red and orange curves are corresponding spline fits to better highlight the fast oscillation fringes. 
    }
    \label{fig:bl40_zeroeig}
\end{figure*}

This daily response to the Sun signal is relatively constant over a period of a year.  Fig.~\ref{fig:beam_vs_theta} shows the directive gain of the dish antennas as computed by an electromagnetic simulation.  The Sun enters the sidelobes of the antennas in a range of polar angles for which the beam patterns are relatively flat.  The simulation is consistent with measurements of the daytime visibilities at different times of the year.    Using the eigenvalue analysis described below, Fig.~\ref{fig:bl40_zeroeig} shows the contribution by the Sun to the visibility for a typical baseline during January, 2018 and then again in April, 2019. As the paths of the Sun through the sidelobes of the antennas are different, at different time of the year, the visibilities are also slightly different. However, we can see that the amplitudes are within $\sim30\%$ of each other. 
% Fig.~\ref{fig:bl40_zeroeig} shows the magnitude of the visibility for a particular baseline for two different periods of time: one from January, 2018 and the other from April, 2019. 

\begin{figure*}
    \centering
    \includegraphics[width=0.49\textwidth,trim = 0 .1 1 1, clip]{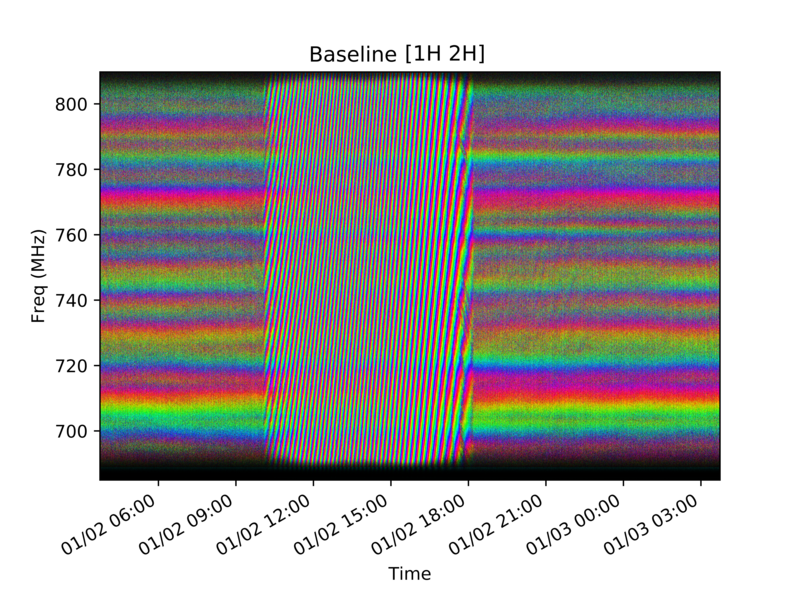}
    \includegraphics[width=0.49\textwidth,trim = 0 .1 1 1, clip]{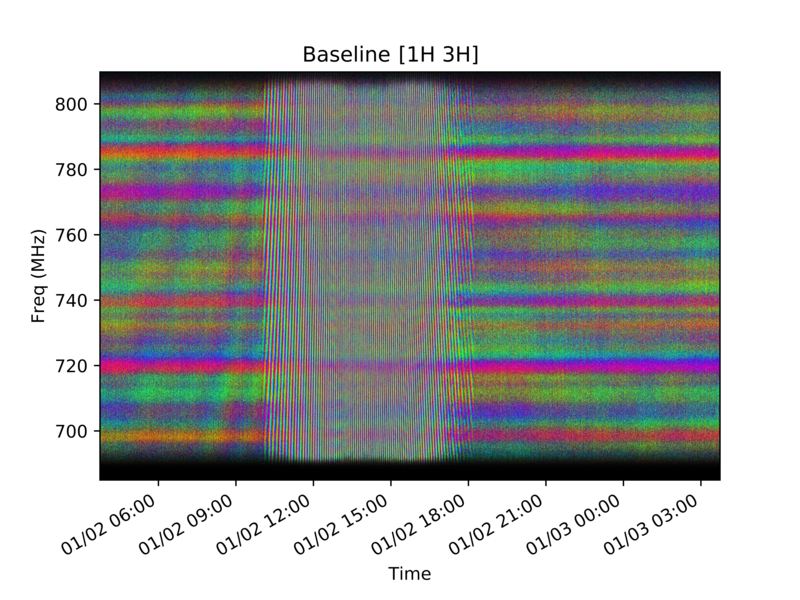}
    \includegraphics[width=0.49\textwidth,trim = 0 .1 1 1, clip]{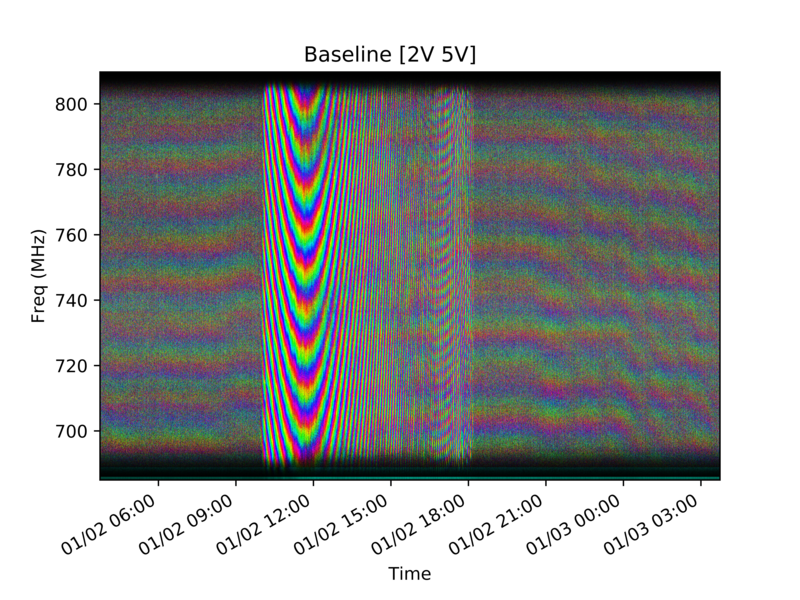}
    \includegraphics[width=0.49\textwidth,trim = 0 .1 1 1, clip]{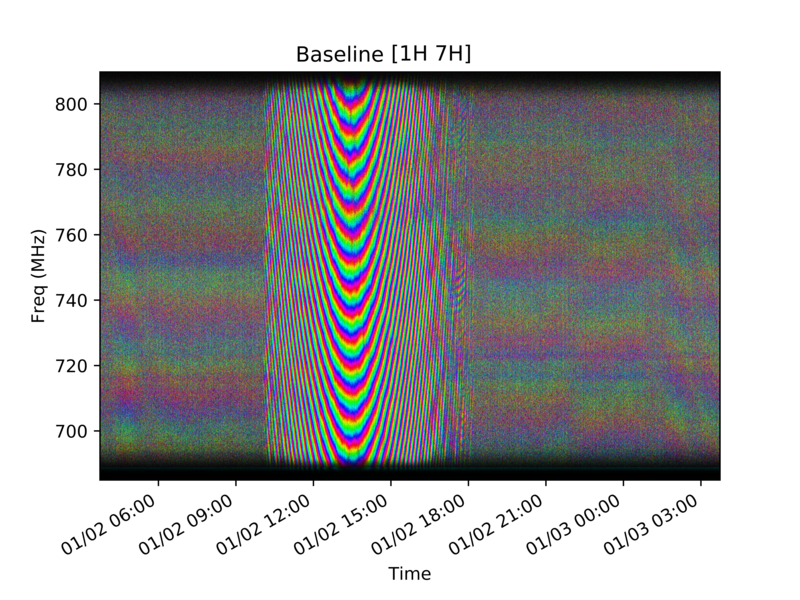}
    \caption{
    Left: The complex visibility for four typical baselines plotted over a 24 hour period.  We represent the phase of the complex visibility by hue (color) and the amplitude by value (brightness) in a HSV (hue, saturation, value) display of the color model (see Fig.~\ref{fig:palette} for details).  In each plot, the local sidereal time proceeds linearly from left to right, with a sampling interval of 1~s. The frequency increases linearly from bottom (685 MHz) to top (810 MHz) in 512 equally spaced frequency bins. The time interval %\rzorange{time interval} 
    from about 10:00 to 18:00 is dominated by the Sun.
    }
    \label{fig:bl40orig}
\end{figure*}

The complex visibilities for four randomly chosen baselines are shown as `waterfall plots' in Fig.~\ref{fig:bl40orig} for a 24 hour period. %{\color{red} We need to plot the color pallet in the appendix. -- DONE}). 
We can see that the daytime data is %much brighter because of the Sun signal. We can  see 
dominated by bright fringes caused by the Sun. On the other hand, the pattern in the nighttime data comes from the much dimmer radio sky and has a very different character. The dominant fringes in the nighttime data come from a combination of weak sources near the NCP and bright sources far from the NCP, particularly Cassiopeia A (Cas A) and Cygnus A (Cyg A). % We can see different fringe rates superimposed on top of each other during the nighttime. 

\section{Removing Sun contamination using eigenvalue analysis}
%\rzorange{
%\begin{itemize}
%\item \st{I suggest to use notations as close as possible to the Tianlai dish paper ($V_{a,b}$) - Use bold-face $\mathbf{V}(t,\nu)$ to denote time and frequency dependent visibility matrix. }
%\item \st{I suggest to remove section 3.3 and and say simply that we are here considering non polarised sky signals, so we don't use cross polarisation visibilities and treat (H,V) polarisations independently (two 16x16 independent visibility matrices).}
%\item \st{I suggest also to start by presenting the visibility matrix structure, as you have done, followed by the expression of the eigen-decomposition of the matrix (what is currently in the beginning of the current subsection 3.4.  
%Maybe, you can use the formulation I showed in one of Tianlai analysis teleconf, which shows clearly that association between the largest, single eigenvalue of the matrix, where there is only one source in the sky, and the noise contribution (I'll send you the slide by e-mail). You continue then with the current section 3.1 (Issues with the autocorrelation signal)} 
%\end{itemize}
%} 

We start by defining the notation used in this paper. The visibility matrix is given by 
\begin{equation}
\mathbf{V}=\left[\mathbf{D}^{\mathbf{s}} \mathbf{G}\right]^{\dagger}\left[\mathbf{D}^{\mathbf{s}} \mathbf{G}\right]+\left\langle[\mathbf{N}]^{\dagger}[\mathbf{N}]\right\rangle,
\end{equation}
\noindent where any individual component is given by 
\begin{eqnarray}
\mathbf{V}_{(i,j)} &=&\left\langle E_{i}^{*} E_{j}\right\rangle.
\end{eqnarray}
\noindent $E_i$ represents the voltage from receiver $i$ and is given by 
\begin{eqnarray}
E_{i} &=&\left(\sum_{s}D_{i}\left(\vec{\omega}_{s}\right) e^{i \mathbf{k}\cdot\mathbf{r}_{i}} F_{s}\right) G_{i} + N_{i},
\end{eqnarray}

\noindent where $F_s$ is the electric field of the radio wave coming
from a source on the celestial sphere, $D_i(\vec{\omega}_{s})$ is the
primary beam of antenna $i$, and this is a function of the direction vector $\vec{\omega}_{s}$. $\mathbf{k}$ is the 3-dimensional wavenumber Fourier dual to the position vector $\mathbf{r}_i$ of feed $i$.
$G_i$ is a direction-independent complex gain factor. $N_i$ is the noise in receiver $i$. 

The intensity of the source at any frequency $\nu$, is given by
\begin{equation}
I_{s}(\nu)=\left|F_{s}(\nu)\right|^{2}=F_{s}^{*}(\nu) F_{s}(\nu)\,.
\end{equation}

For extended sources we need to integrate over different directions for calculating $E_i$: 

\begin{eqnarray}
E_{i} &=&\left(\int D_{i}\left(\vec{\omega}_{s}\right) e^{i \mathbf{k}\cdot \vec{r}_{i}} F_{s} d\omega_s \right) G_{i} + N_{i}\,.
\end{eqnarray}

The visibility is an ensemble average of the $E_i^*E_j$, i.e.
%Therefore, we have visibility data integrated over the integration time period, i.e. 
%During daytime, the largest contribution from the sky in the Sun 
\begin{eqnarray}
    \mathbf{V}_{(i,j)}&=&\langle \,E^*_{i}\,E_{j}\, \rangle_{\tau_\mathrm{int}}  \nonumber \\
    &=&  \Big[\,\frac{1}{\tau_\mathrm{int}} \int_0^{\tau_\mathrm{int}} E^*_i \,E_j\;\mathrm{d}t\;\Big]
\end{eqnarray}

\noindent where $\tau_\mathrm{int}$ is the integration time, which is constant for any time and frequency bin $(t,\nu)$. For the current Tianlai setup, the integration time is $1$~s. The asterisk $(\;*\;)$ represents the complex conjugate  and the bracket $\langle \;\;\rangle$ represents the ensemble average.

Here we should note that the visibilities from different astrophysical sources are additive. 
Provided there is only one point source on the sky the visibility matrix, i.e. $\mathbf{V}_{(i,j)}$ at any time can be written as an outer product, of the electric field from the source measured at different feed antennas.
Therefore, if the visibility matrix is decomposed into its corresponding eigenvalues and eigenvectors, there should be only one nonzero eigenvalue.
In the presence of other weaker sources,
%which are not correlated with the strongest one, 
%\rzorange{\st{What do you mean by not correlated. I think we are simply modeling the instantaneous visibility matrix as due to one or more sources in the sky, plus noise, and we assume that the noise is uncorrelated between different feeds.}} 
the largest eigenvalue should correspond to the Sun signal and the eigenvector corresponding to the largest eigenvalue will roughly point toward the direction of that source in the eigenspace. The contributions from additional, weaker sources and noise may alter the direction slightly.

%Though, in such cases, 
% roughly the direction of the Eigen vector will correspond to the strongest source. % provided the source is strong enough than the other sources. 

%Theoretically, if we have the product of the electric fields from a source detected by two antenna then we can decompose them in the Eigen vectors and seperted of the signa. 

Comparing the visibility amplitudes between the daytime and the nighttime data in Fig.~\ref{fig:Fig1}, we can infer that the largest contribution to the daytime signal is from the Sun, entering through the antenna sidelobes. Therefore, in the eigen decomposition of the visibility matrix the largest eigen value should represent the solar contamination. 
%Provided that the integration time is small compared to the Sun fluctuation during this time, we can safely assume that the signal from the Sun and other sources  remain constant during this period. 
%Therefore in the eigen value decomposition of the visibility matrix we should get the 

%\rzorange{\st{One can go to significantly higher integration time , about a minute, as long as the sky rotation stays small compared to the array resolution during the integration. The sources in the sky are indeed uncorrelated, unless at very small angular distances. This is a main hypothesis that makes radio interferometry possible.} } %As the Sun is the strongest source in the sky during the day, the largest eigenvalue should correspond to the Sun signal. 

\begin{figure*}
  \centering
   \includegraphics[width=.47\textwidth]{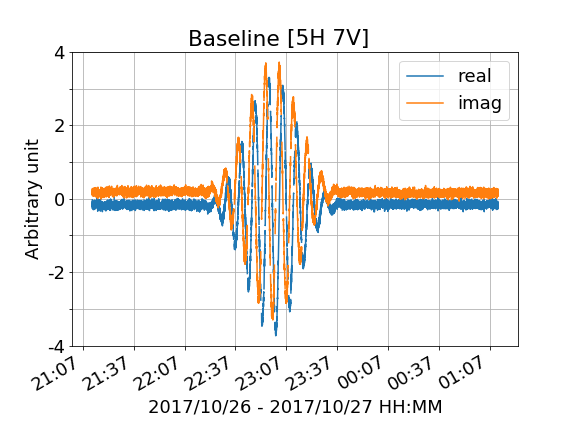}   \includegraphics[width=.47\textwidth]{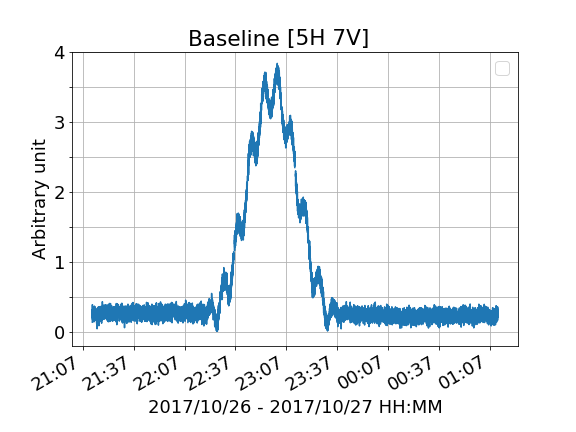}

%\end{figure*}
%\begin{figure*}
  \centering
    \includegraphics[width=.47\textwidth]{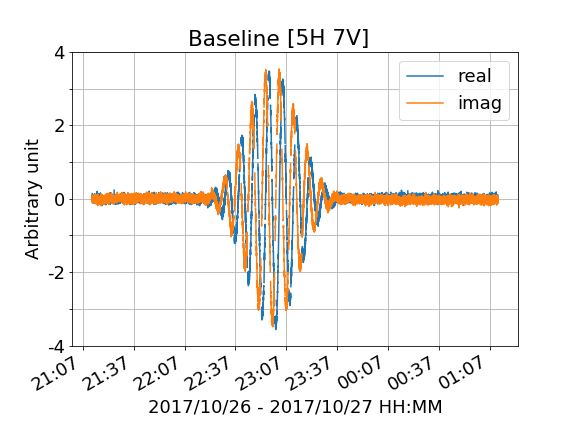}
   \includegraphics[width=.47\textwidth]{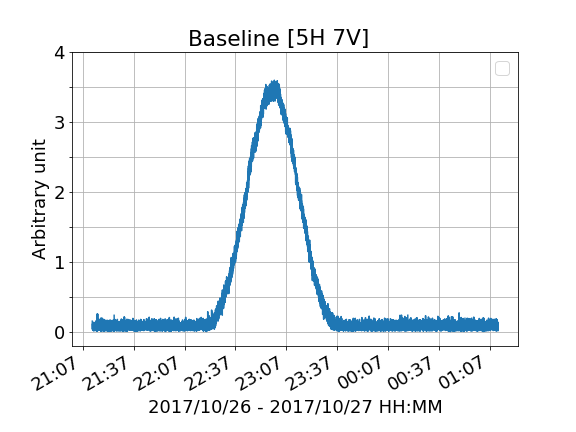}
\caption{ %\footnotesize 
Top-Left: The real and imaginary components of the raw visibility of baseline [5H 7V] during transit of Cas A in October 2017. We can see that there is a small DC offset in both the real and imaginary components. Top-Right: Amplitude of the visibility of baseline [5H 7V] before removing the DC offset. We can see that there is an oscillatory pattern on top of the Gaussian transit peak.%}
%\label{fig:bl45RI}   
%\caption{ %\footnotesize 
Bottom-Left: The real and imaginary components of the raw visibility after removing the offset form each of the %the real and imaginary 
components. Bottom-Right: Amplitude of the visibility of baseline [5H 7V] after removing the mean. We can see a perfectly  Gaussian transit peak. %\rzorange{\st{Maybe we can combine Fig 7 and Fig 8 into a single figure, top panel, before DC offset subtraction, bottom panel, after DC offset subtraction }} % is recovered.
}
\label{fig:bl45amp}
\end{figure*}

%In eigen-decomposition of the visibility matrix, the largest eigenvalue during the daytime represents signal corresponding to the Sun, and the corresponding eigenvector represents the direction of the Sun in the eigenspace.

\subsection{Issues with the autocorrelation signal}

The voltage from the feeds contain a contribution from the receiver noise. %cross feed coupling etc. %If 
%the instrumental noise from feed $i$ be $N_i$, 
Therefore, the %\rzorange{}
measured signal or voltage $E_i$ for a given feed $i$  is the sum of the sky signal, $E_{\text{Sky}\,i}$  and the instrument noise, $N_i$, i.e.  $E_i=E_{\text{Sky}\,i}+N_i$.

Under the assumption that the noise terms from separate feeds are uncorrelated, we can say that the ensemble average of the noise from feed $i$ and feed $j$ is zero, i.e. $\langle N_i^* N_j\rangle\approx 0$. Therefore, the visibility for cross-correlated feed $i$ and $j$, where $i\neq j$, is $\mathbf{V}_{(i,j)} \approx \langle E_{\text{Sky}\,i}^{*} E_{\text{Sky}\,j} \rangle$.

However, for the autocorrelations, the visibilities, $V_{(i,i)}$ are dominated by the positive noise term $\langle N_i^* N_i\rangle$. The amplitudes of the autocorrelation signals are much higher than those of the cross-correlation signals. Therefore, in an eigen-decomposition of the visibility matrix, the eigenvectors are dominated by the noise signals from the autocorrelation, as the sky signals are typically much smaller than the noise. 

It is not possible to ignore these auto-correlation signals or simply set them to $0$ during the eigenvalue decomposition. To overcome this difficulty, we replace the corresponding terms in the visibility matrix by the following quantity as a proxy for the autocorrelation visibilities. 

\begin{equation}
%\label{autocorr}
    \mathbf{V}_{(i,i)}=\frac{1}{n}\sum_{k,j}\mathrm{abs}\left[\frac{\mathbf{V}_{(i,k)}\mathbf{V}_{(j,i)}}{\mathbf{V}_{(j,k)}}\right]\;,\hspace{.3cm}\forall i\ne j \ne k
    \label{Eq.autocorrelationRenorm}
\end{equation}
Here, $n$ is the number of values over which we are doing the sum, i.e. the number of $(j,k)$ pairs. This brings the level of the amplitude of the autocorrelation to the order of the cross-correlation amplitude and we can do a meaningful eigenvalue decomposition. 

\begin{figure*}
    \centering
    \includegraphics[width=0.49\textwidth,trim = 0 1 1 1, clip]{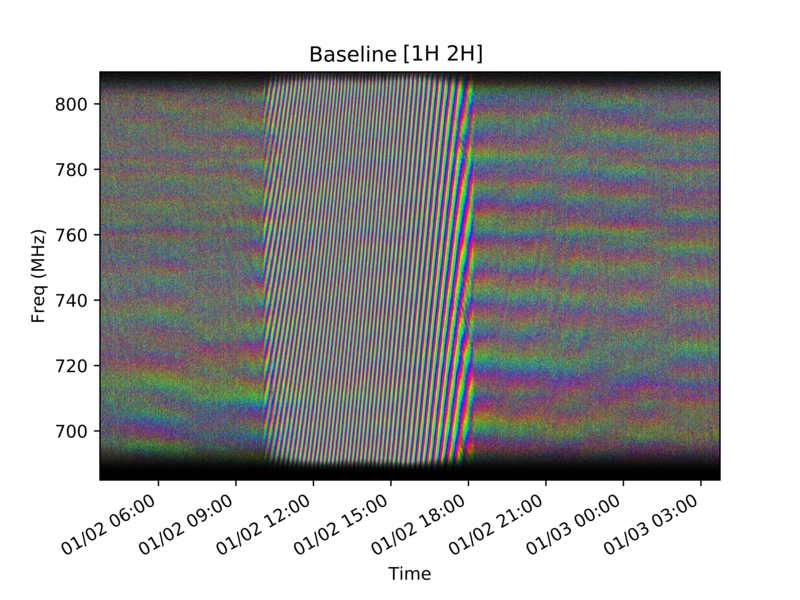}
    \includegraphics[width=0.49\textwidth,trim = 0 1 1 1, clip]{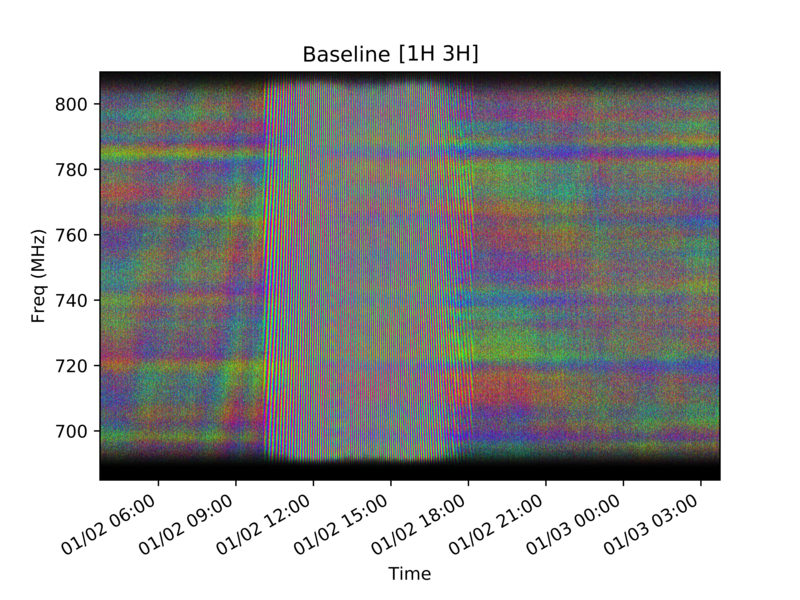}
    \includegraphics[width=0.49\textwidth,trim = 0 1 1 1, clip]{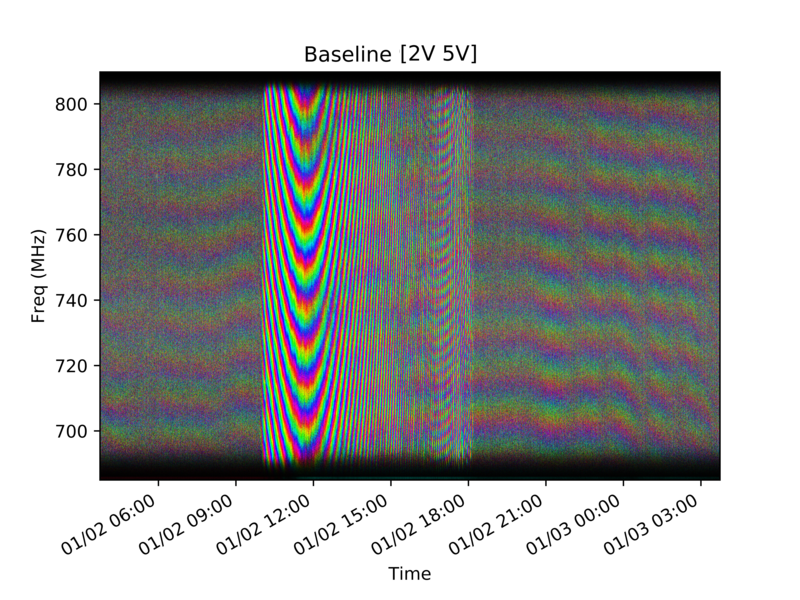}
    \includegraphics[width=0.49\textwidth,trim = 0 1 1 1, clip]{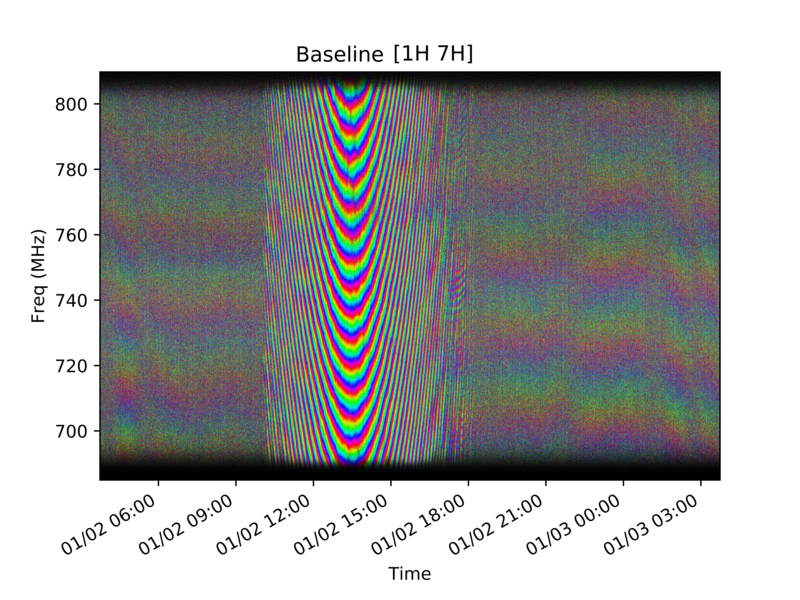}
    \caption{
    The waterfall plot of the complex visibility (same as Fig.~\ref{fig:bl40orig}) after the nightly mean subtraction. We can see that most of the horizontal stripes, which probably are caused by cross talk, are now gone from the waterfall plot. The structures from the sky are more prominently visible.
    }
    \label{fig:bl40mr}
\end{figure*}

\subsection{DC offset in the visibility}

If there is no strong source in the sky then the real and the imaginary parts of the visibility are expected to randomly fluctuate around $0$.  However, often in radio interferometers, there are some DC offsets in the real and imaginary components of the visibility. 
The offsets may originate from a variety of 
%{\color{red}  from the cross-talk between the dishes} or 
 systematic effects, and cross-coupling of signals between the antennas is one of them. In the Tianlai data, we see it in multiple baselines as colored horizontal stripes in the waterfall plots of the complex visibility (see Fig.~\ref{fig:bl40orig}). %We believe the dominant mechanism for this cross-coupling is transmission of receiver noise between antennas. 
 
%As an example of the need to remove this mean, 
In the top-left plot of Fig.~\ref{fig:bl45amp}, we show the real and the imaginary parts of the visibility from a transit of Cas A observed by baseline [5H 7V].  On the top-right of the same plot, we show the amplitude of the visibility during a transit of Cas A, which is %The amplitude of the visibility during the transit is 
expected to form a Gaussian profile. However, the plot shows a wavy feature modulating the Gaussian. This pattern comes from the DC offsets in the real and the imaginary parts of the visibility, shown in the left plot.  At the beginning of the plot, when there is no source, we can still see some DC signal in the real and the imaginary part and they do not fall on top of each other. %This offset is actually responsible for the wavy pattern in the absolute value of the visibility. 
% as shown in the right of the same plot.

In the bottom panel of Fig.~\ref{fig:bl45amp}, we show the amplitude, as well as the real and the imaginary component of the visibility after subtracting the mean of the nighttime data from both the real and imaginary components of the visibility for each frequency channel and each baseline.  The amplitude of the visibility, after DC offset removal, shows a Gaussian peak during the transit of Cyg A, as expected.
%Therefore, removing the DC offset from the visibility is another important thing that should be removed before any analysis. 

 The presence of this DC offset may also introduce an error in the eigen-decomposition and it must be removed before running the Sun removal algorithm described below.
 We subtract the mean value of the real and imaginary parts of the visibility for each night of data. We do not include the daytime data when computing the mean, because it is contaminated by the Sun. However, the DC offset is very stable over each night and from night to night. So we remove the nightly mean from the entire 24 hours of data, including the daytime data. This removal also reduces the night-to-night variation both in absolute terms and as a fraction of the remaining signal~\citep{wu2020tianlai}.

%If there is no DC offset, we expect that the average value of the real or the imaginary part of the visibility will be $0$. However, this is not the case as seen from the figure; both the real and the imaginary components have some DC offset. %The origin of this offset may have been from the cross-talk between the feeds. 

\begin{figure*}
    \centering
    \includegraphics[width=0.9\textwidth]{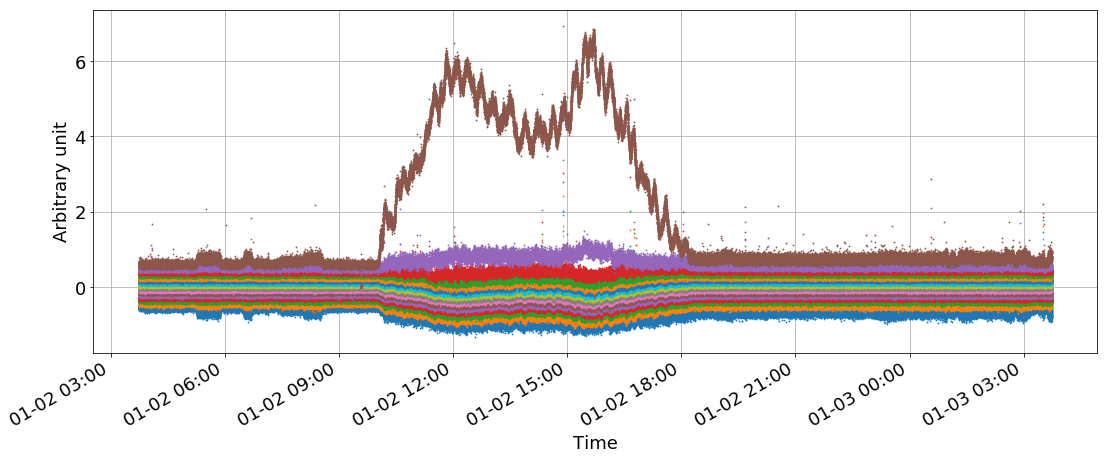}
    \caption{Plot of all the 16 eigenvalues in the eigen-decomposition of the horizontal polarization visibility $\mathbf{V}^{(H)}$. We can see that one of the eigenvalue is much larger than the other eigenvalues during daytime. This particular eigenvalue is coming from the solar contamination of the daytime data. We can see that the other eigenvalues are also affected during daytime. This happens due to the change in the eigenvectors, one of which (the eigenvector corresponding to the largest eigenvalue) is oriented towards the Sun during the daytime. The plot here is shown at the central frequency (747.5 MHz) of the observed Tianlai Dish Array band. All other one dimensional plots also use this frequency.
    }
    \label{fig:eigenvalues}
\end{figure*}

\begin{figure*}
    \centering
    \includegraphics[width=0.9\textwidth]{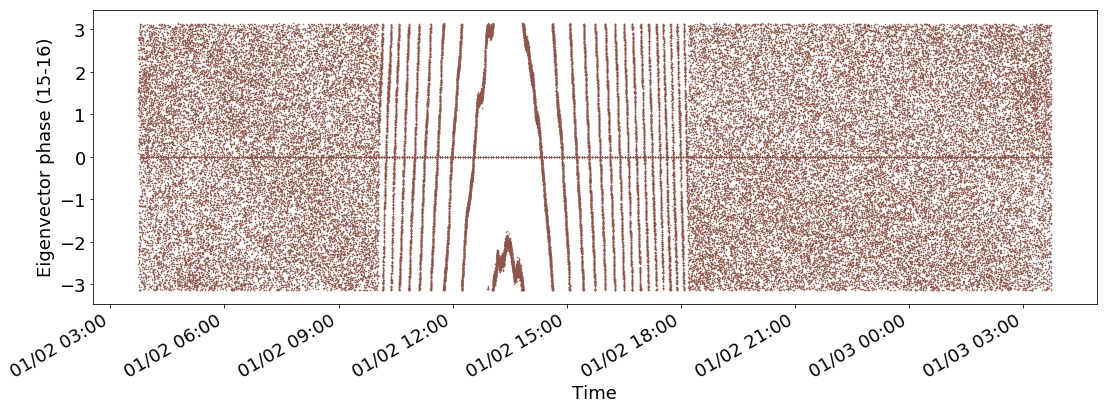}
    \caption{Phase plot of the 15th component of the eigenvector corresponding to the largest eigenvalue of the horizontal polarization visibility, $\mathbf{V}^{(H)}$. We can see the strong fringes during daytime, which confirms that the eigenvalue is  coming from a single strong source, which is the Sun. During night, as there is no single strong source, the phase is varying randomly.  
    The horizontal line in the center is caused by the calibration noise source, which is turned on and off periodically.
    }
    \label{fig:eigenvector1}
\end{figure*}

%{\color{red}Subtraction of the nightly mean removes much of the correlated noise but also a significant fraction of the signal (gain times sky).  Since the sky signal should be the same at the same LST it does not contribute to night-to-night variation which can be due to variations in gain or in correlated noise. One would not expect that subtracting the nightly mean would decrease the fractional variation if the variation were only due to gain fluctuations, so we infer that much of what was subtracted is correlated noise. Nightly mean subtraction remaining signal is predominantly from the sky(from Tianlai Dish paper. Need to cite that paper)} 

In Fig \ref{fig:bl40mr}, we show the waterfall plots of the complex visibilities from four baselines after the nighttime mean removal.  We can see the nighttime structures more prominently after the mean subtraction. %After the nightly mean removal, we can see the the fringe pattern as expected for bright local sources on the sky. - This line should not come here.

%Fig.\ref{fig:bl45RI}, Fig.\ref{fig:bl45amp}

%\subsection{Problem with cross polarization}
%\st{\rzorange{I suggest to remove this subsection \\}}
%Each of the Tianlai dish antennas has a dual polarization feed. If we consider both polarizations in our analysis then theoretically we should get two large eigenvalues during the eigen-decomposition. However, the cross polarization  between the two feeds from the same dish is much higher than the cross polarization signal from different dishes. %The exact reason is not known. However, most likely cause for it is the %it %happens due to some coupling between two antennas in the same dish.  
%This is not surprising, because the two dipoles in the feed are adjacent to each other.
%Just as for the autocorrelation signals, it is not possible to ignore these cross-correlation signals by simply setting them to $0$. %Such a procedure yields completely wrong eigenvalues.

Here, we also like to point out that in this work, for simplicity we have only considered the auto-polarization signals. %Therefore, while decomposing the eigenvalues we will treat each polarization separately, and 
%do not mix the two polarizations. Our 
We set the cross polarization signal to $0$, making the visibility matrix look like 
\begin{equation}
    \mathbf{V} = \begin{bmatrix}
    \mathbf{V}^{(H)} & \mathbf{0}  \\
    \mathbf{0} & \mathbf{V}^{(V)}
\end{bmatrix},
\label{Eq.ZeroCrossPol}
\end{equation}
where the superscripts $H$ and $V$ refer to the horizontal or vertical polarization, respectively. $\mathbf{V}$ is the visibility matrix at each time and frequency $(t,\nu)$. % We set all the cross polarization terms between horizontal and vertical feeds to be zero. 

\subsection{Understanding the eigenvalue decomposition}

For removing the solar contamination from the real data, we first remove the DC offset from all the cross-correlation channels. We write the visibility matrix for each frequency and at every time bin in the form shown in Eq.~(\ref{Eq.ZeroCrossPol}). We replace the autocorrelation signals using the formula given in Eq.~(\ref{Eq.autocorrelationRenorm}) and decompose the matrix into eigenvalues and eigenvectors as $\mathbf{V} = \mathbfcal{E} \mathbf{\Lambda} \mathbfcal{E}^{-1}$.
At this point it should be noted that the eigenvalue decomposition is invariant under a U(1) transformation, i.e. if we multiply the full eigenvector matrix, $\mathbfcal{E}$, by a factor of $e^{i\psi}$ for any real $\psi$, then the corresponding eigenvalue matrix $\mathbf{\Lambda}$ will remain invariant. Therefore, without loss of generality, we choose the first component of the eigenvector for each time and frequency component to be real and positive.

%\rzorange{I am not sure, I think that you can indeed apply a overall phase factor to the eigenvector matrix, so you should be able to have the first component of {\bf ONE} of the eigenvectors real and postive, but not all of them. In my opinion, you can not apply a differnt phase factor to each eigenvector independently. }
%the eigen-decomposition is invariant under $U(1)$ transformation. To avoid that ambiguity we set the first component of the eigen vector correspond to the largest eigen value to real and positive. 

Also, in our case the visibility, $\mathbf{V}^{(X)}$  (where $X=\{H,V\}$) is a block diagonal matrix. 
Therefore, the eigenvalues and the eigenvectors of the matrix will be the eigenvalues and eigenvectors from each of the blocks, i.e. 
$\mathbf{V}^{(X)} = \mathbfcal{E}^{(X)} \mathbf{\Lambda}^{(X)} \mathbfcal{E}^{(X)\,-1}$. 
For each $t$ and $\nu$, $\mathbfcal{E}$ is a $n\times n$ matrix whose whose $i$-th column is the complex normalized eigenvector, $\mathcal{E}^{(X)}_{i}$ of $\mathbf{V}$, and $\mathbf{\Lambda}$ is the diagonal matrix whose diagonal elements, $\mathbf{\Lambda}_{ii} = \lambda_i$, are the corresponding eigenvalues.

%A plot of the visibility after solar contamination removal for four typical baselines is shown in In Fig.~\ref{fig:bl40orig}. We plot the phase of the complex visibility by hue (color) and amplitude by value (brightness) in a HSV (hue, saturation, value) representations of the color model. The color representation is specified in Appendix ?. We can see that the daytime data is much brighter because of the Sun signal. We can also see the fringes in the daytime data. On the other hand, the pattern in the nighttime data is coming from the sky and is much different. Different fringes in the nighttime data are coming from different sources in the sky and is more frequency-dependent. We can see different fringe rates superimposed on top of each other during the nighttime. 

\begin{figure*}
    \centering
    \includegraphics[width=0.9\textwidth]{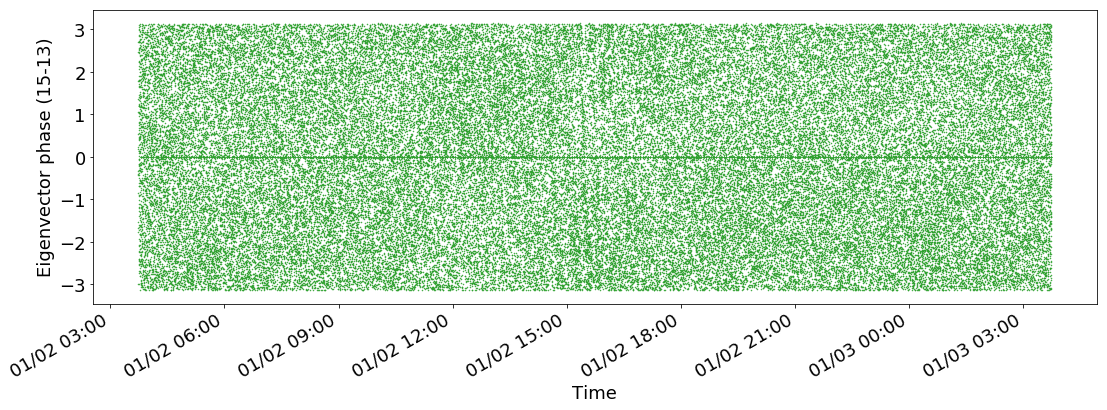}
    \caption{Phase plot of the 15th component of the eigenvector corresponding to the 4th largest eigenvalue in the horizontal polarization visibility $\mathbf{V}^{(H)}$. Here we don't see any fringes, showing that no individual strong source is contributing to this particular eigenvalue.  %\rzorange{Indeed, with 1s integration, the visibilities are noise dominated, except when there is a very strong bright source in the sky (CasA, or the sun). What do we get if we look at the first few (1,2,3) largest eigenvalues in night time, with around a minute integration time ? }
    }
    \label{fig:eigenvector2}
\end{figure*}

%\begin{figure*}
%    \centering
%    \includegraphics[width=0.9\textwidth]{figs/eigenphase1513_2.png}
%    \caption{1 min average of the above plot
%    }
%    \label{fig:eigenvector2}
%\end{figure*}

\begin{figure*}
    \centering
    \includegraphics[width=0.98\textwidth]{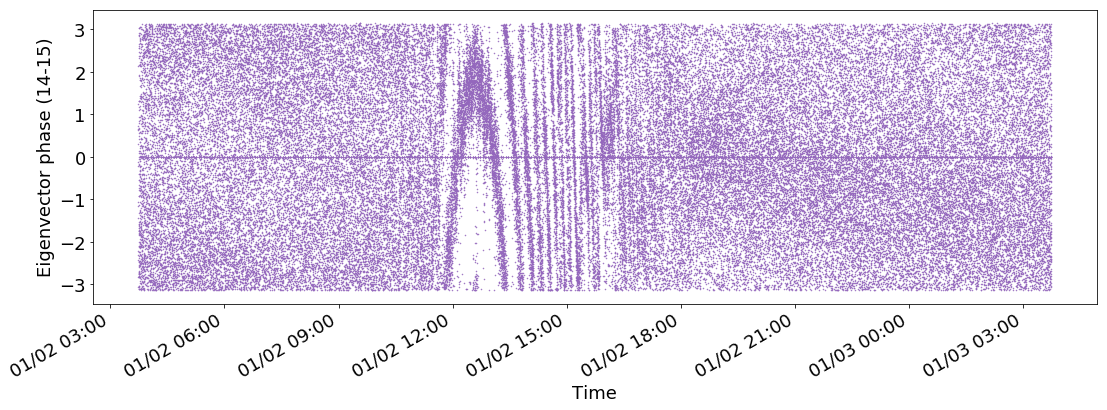}
    \caption{The phase of one component of the eigenvector corresponding to the 2nd largest eigenvalue. We can see weak fringes, indicating that some of the solar contribution is present in the second largest eigenvalue. 
    }
    \label{fig:eigen_vector_secondlargest}
\end{figure*}

%\begin{figure*}
%    \centering
%    \includegraphics[width=0.98\textwidth]{figs/eigenphase1415_2.png}
%    \caption{1 min average of the above plot 
%    }
%    \label{fig:eigen_vector_secondlargest}
%\end{figure*}

Fig.~\ref{fig:eigenvalues} shows $16$ eigenvalues calculated from the horizontal polarization matrix ($\mathbf{V}^{(H)}$) 
%\rzorange{}
as a function of time. The plot clearly shows that one eigenvalue is much higher then the other values during daytime. We can undoubtedly infer that the major contribution to the power in that particular eigenvalue comes from the solar contamination, as the Sun is by far the strongest source in the sky during day time. 

Fig.~\ref{fig:eigenvector1} shows the phase of one of the components of the eigenvector that corresponds to the largest eigenvalue:  $\mathbfcal{E}^{(X)}_{(15,16)}$.
%\rzorange{Should'nt this be $\mathbfcal{E}^{(X)}_{(15,16)}$ (15 instead of 14 ?) }{\color{blue} I think there is a typo.Fixed.}
Eigenvalues are sorted %\rzorange{sorted [instead of organized ?]}
according to the daytime amplitude. The 16th eigenvalue is the largest and we have shown the 15th component of the corresponding eigenvector. 

%\rzorange{\st{As I said before, I think that it would be better to put matrix row,column indices on the right, to stay with the usual convention $V(i,j)$ or $V_{t,\nu}(i,j)$. By the way, in the text, you say 15 th component, so, it should be (15,16) I guess. I think also that it would be better to sort eigenvalues in descending order, so that the largest eigenvalue becomes the first. That makes the rank of the largest eigenvalue independent of the array size.}}

Clear fringes are visible in the daytime data, showing that the daytime signal in the eigenvector is coming from a single strong source. In the nighttime data we can see the phase varies completely randomly, proving the absence of any single strong source at the nighttime data.

In Fig.~\ref{fig:eigenvector2} we show the phase from one of the components of the 4th largest eigenvector, $\mathbfcal{E}^{(X)}_{(15,13)}$. Unlike Fig.~\ref{fig:eigenvector1}, no fringes are visible, indicating that there is no single strong source being detected by that particular baseline and the signal is coming from the background sky. The same thing is true for any of the other smaller eigenvalues. In Fig.~\ref{fig:eigen_vector_secondlargest} we have plotted the phase from one component of the eigenvector corresponding to the second largest eigenvalue. We can find weak fringes during the daytime, indicating that some of the Sun signal has `leaked' into this eigenvector. Ideally, the second eigenvalue represents the second strongest sources in  the sky, and this leakage may be due to the presence of other sources and the background noise. In addition, the re-normalization of the autocorrelation signal using Eq.~\ref{Eq.autocorrelationRenorm} is another possible cause of this leakage. Finally, it may also be that some of the solar radiation is being reflected from the ground and illuminating the feeds from a different direction from the main Sun signal.

%{\color{red} Rewrite}
%Whatever be the pattern of the beam, as the signal is coming from the same direction of the sky, theoretically we can use an eigenvalue analysis to remove the signal from the Sun. However, in practice, the antenna gain fluctuation, autocorrelation noise, side-lobe gain pattern, ground reflection, and crosstalk between antennas introduce mixing between the largest eigenvector and other smaller eigenvectors. Therefore, singling out and removing the Sun signal is not straightforward. 

If we plot the phase from any component of the eigenvector corresponding to the third largest eigenvalue, we can still see some fringe pattern in the daytime data. However, these fringes are much weaker in comparison to $\mathbfcal{E}^{(X)}_{(15,x)}$ showing that the leakage of solar power is mostly restricted to the second largest eigenvalue. 

\begin{figure*}
    \centering
    \includegraphics[width=0.9\textwidth]{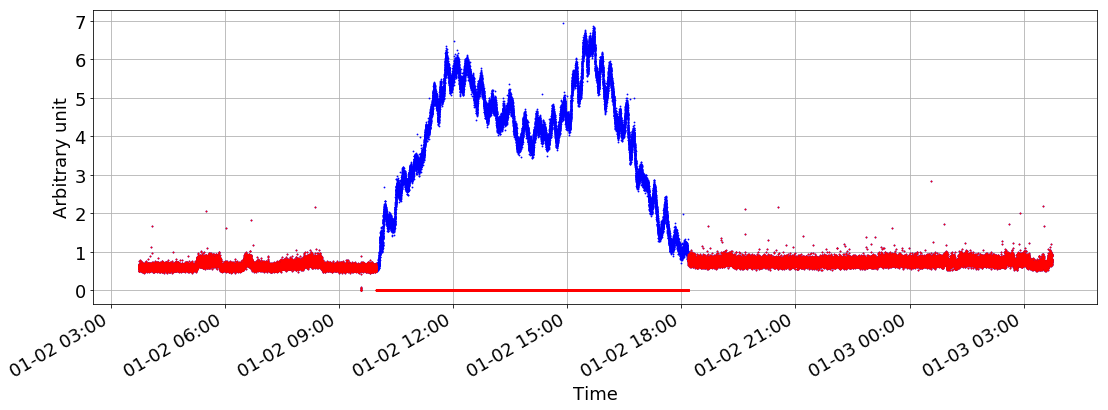}
    \caption{Blue: The original largest eigenvalue from the eigen-decomposition, during daytime.  Red: The same largest eigenvalue but with zero value during the daytime. This step simulates removing the Sun signal in Eq. \ref{rm_Sun}, since the largest eigenvector during the daytime points in the direction of the Sun in the eigenspace.
    }
    \label{fig:zero_eig}
\end{figure*}

\subsection{A first attempt to subtract the solar contamination signal}

%\rzorange{\st{Maybe the title should be : A first attempt to subtract the solar signal ?}}

\begin{figure*}
    \centering
    \includegraphics[width=0.49\textwidth,trim = 0 1 1 1, clip]{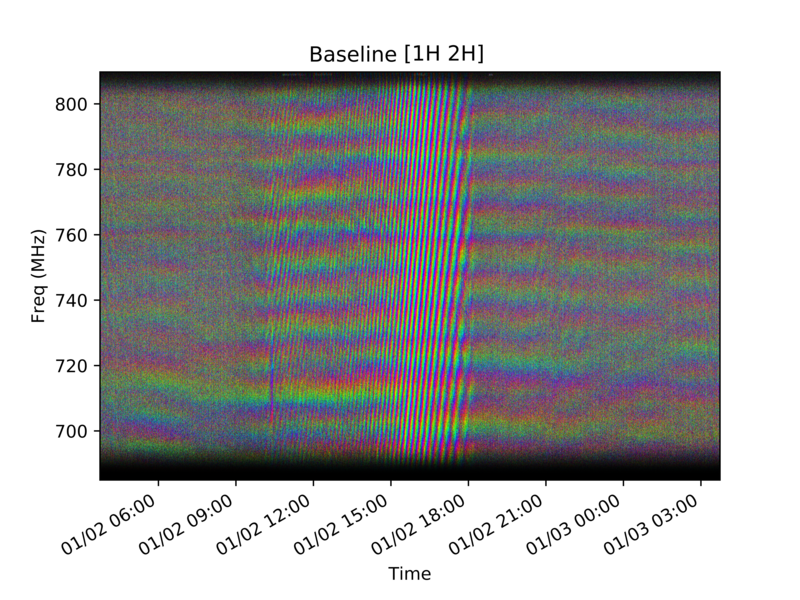}
    \includegraphics[width=0.49\textwidth,trim = 0 1 1 1, clip]{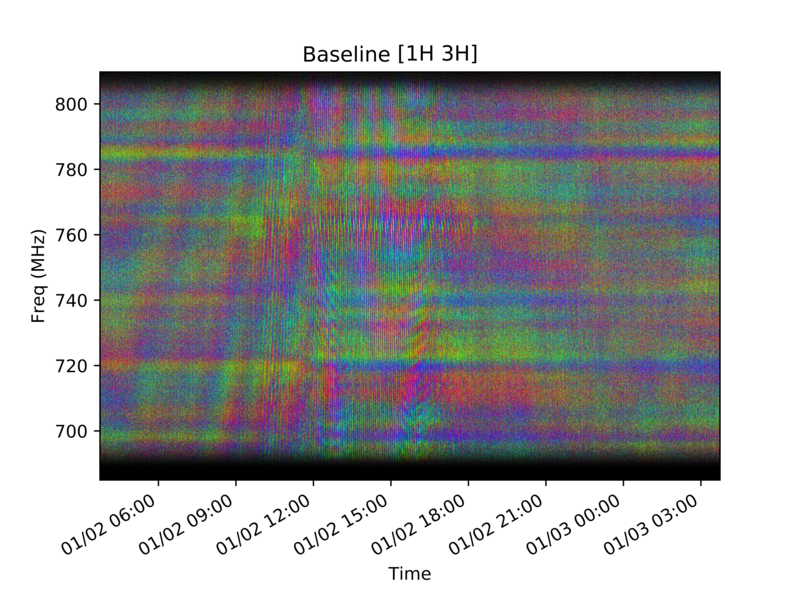}
    \includegraphics[width=0.49\textwidth,trim = 0 1 1 1, clip]{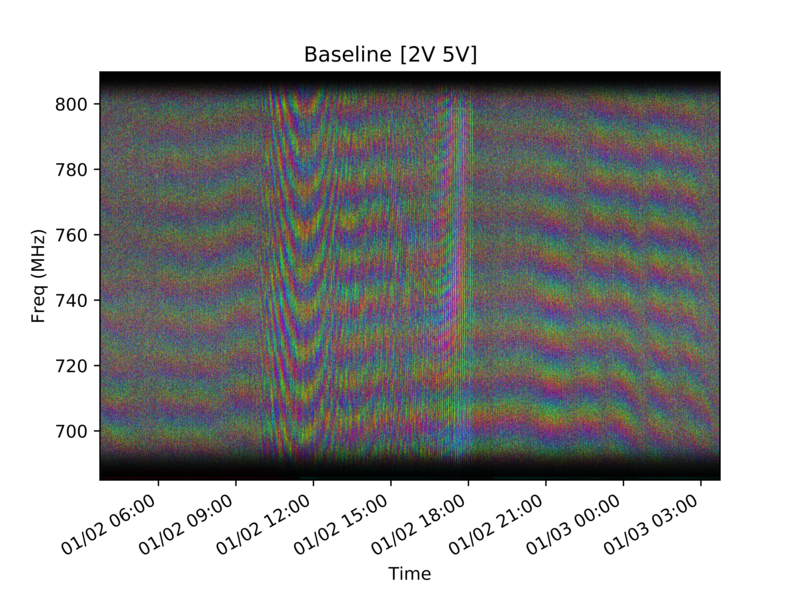}
    \includegraphics[width=0.49\textwidth,trim = 0 1 1 1, clip]{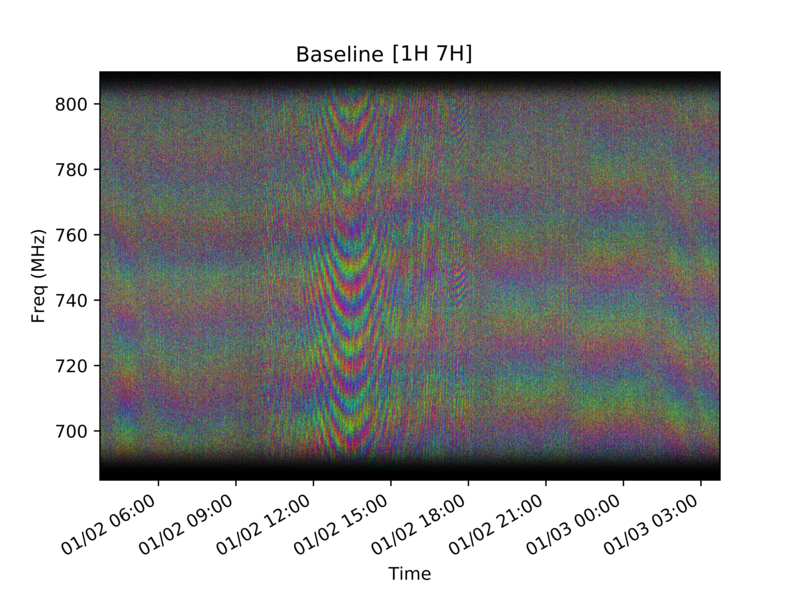}
    \caption{
    %Left: 
    The waterfall plot of the complex visibility after zeroing the largest eigenvalue during daytime for four typical baselines. Nightly mean subtraction has been applied. There is still some residual solar contamination signal, which is causing the weak fringes during the daytime.
    }
    \label{fig:zero_eigseparation}
\end{figure*}

Because the solar signal is contributing mostly to the largest eigenvalue, as a first step in removing the Sun signal we can set the largest eigenvalue during the daytime data to $0$, and then reconstruct the visibility. In Fig.~\ref{fig:zero_eig} we show the largest eigenvalue as a function of time (in blue during the daytime). The red curve shows the value after setting the largest eigenvalue during daytime to be $0$. All the other eigenvalues are kept fixed. In Fig.~\ref{fig:zero_eigseparation}, we show the waterfall plot of complex visibilities from 4 baselines that we recover after this step. 
The plots show that most of the contamination is removed. However, some solar contamination signal is still discernible in the visibility plot. We can see clear, faint fringes for each of the baselines plotted in Fig.~\ref{fig:zero_eigseparation}. 

%After the normalization step, we assume that the largest eigenvalue in an eigen-decomposition of the visibility matrix comes from the Sun signal. To remove the Sun signal from the data, we do an eigen-decomposition of the complex visibility matrix at each time and frequency.

%We set the largest eigenvalue to zero and reconstruct the visibility matrix. The reconstructed visibility matrix will have the contamination from the Sun removed.

%Another possibility of the daytime signal may be the Sunrays reflected from the ground. However the Tianlai data does not show strong contribution from the reflected signal.

%If the Sun signal is the dominant signal in the sky, along with other weaker sources, and if the signals from the sources are un-correlated, then if we decompose the autocorrelationa and the cross-correlation visibilities in the eigenvalue and the eigenvector, the height eigenvalue should corresponds to the strongest source in the sky. The eigenvector corresponding to the largest eigenvalue will point in the direction of that source in the eigenspace. An ideal Sun removal technique should filter out this largest eigenvalue while retaining the signals from other sources in the sky. %Work on this project is in progress.

%we can't use the data from the daytime if the Sun contamination is not removed. Therefore in this paper we have use some method which will be able to remove the daytime signals. 

\begin{figure*}
    \centering
    \includegraphics[width=0.9\textwidth]{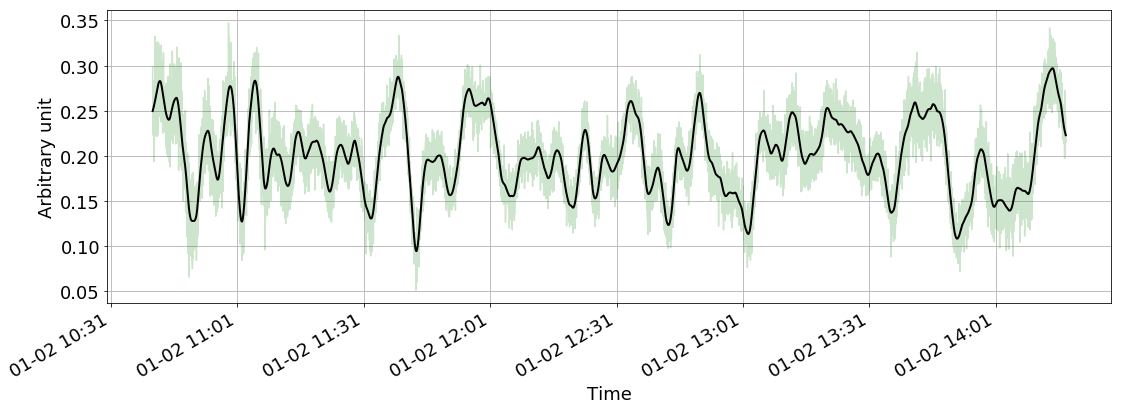}
    \caption{A 4-hour segment of the amplitude of one component of the eigenvector corresponding to the largest eigenvalue is shown in the light green curve. The sampling interval is 1~s. The random fluctuations in the data come from the noise. The black curve shows this component after smoothing the data, as described in section~\ref{sec:smoothing}. 
    }
    \label{fig:eigen_vector_smoothing}
\end{figure*}

\begin{figure*}
    \centering
    \includegraphics[width=0.49\textwidth,trim = 0 .1 1 1, clip]{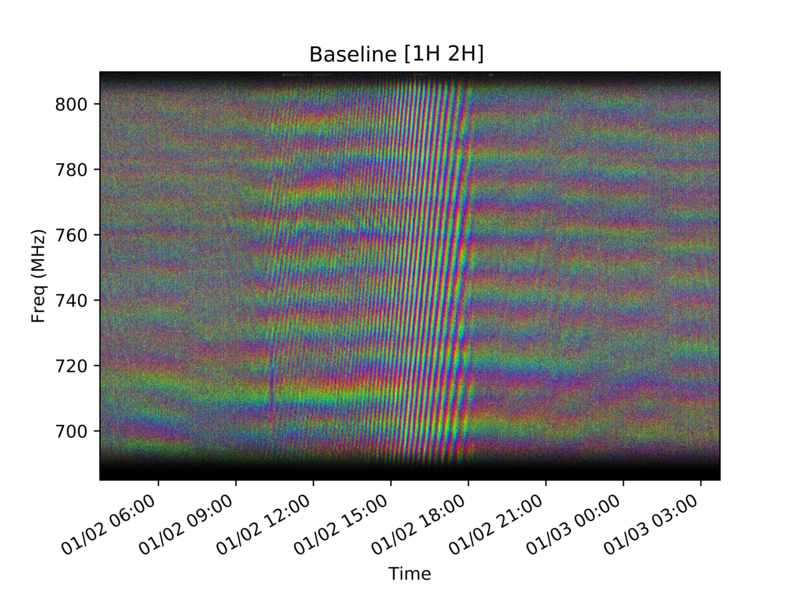}
    \includegraphics[width=0.49\textwidth,trim = 0 .1 1 1, clip]{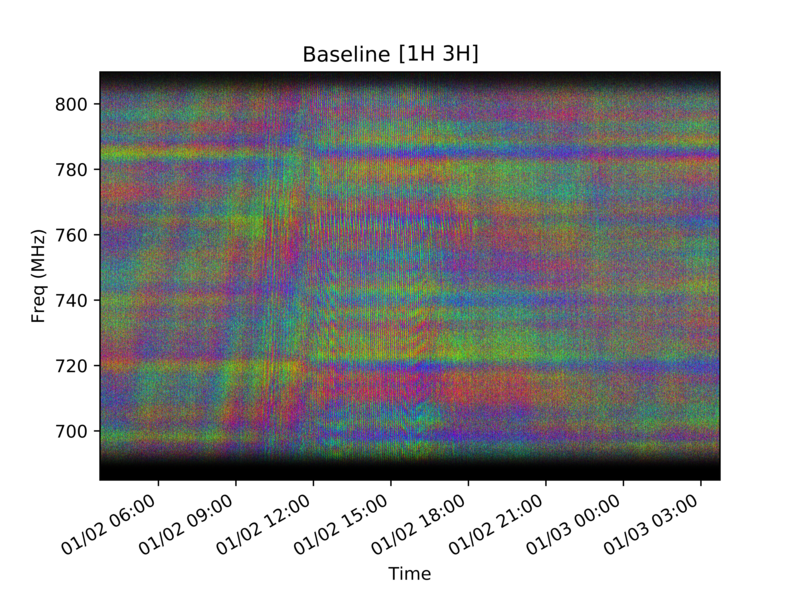}
    \includegraphics[width=0.49\textwidth,trim = 0 .1 1 1, clip]{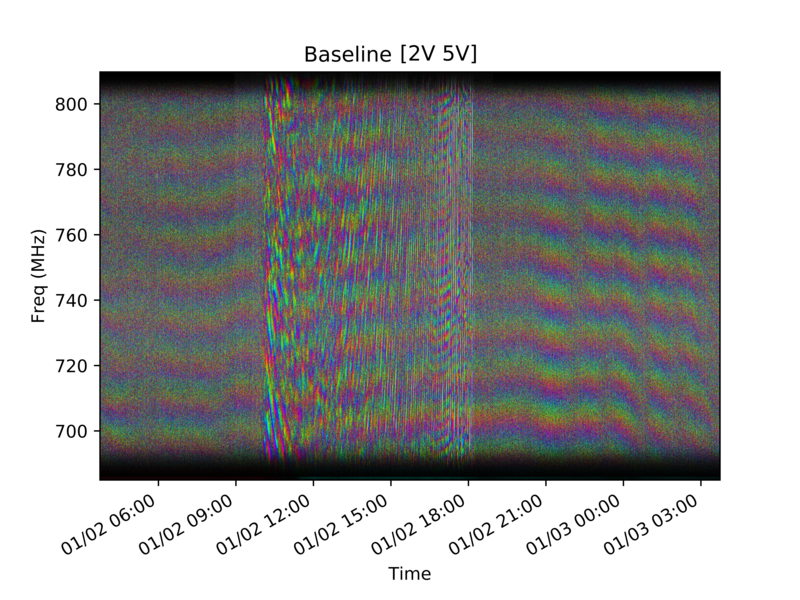}
    \includegraphics[width=0.49\textwidth,trim = 0 .1 1 1, clip]{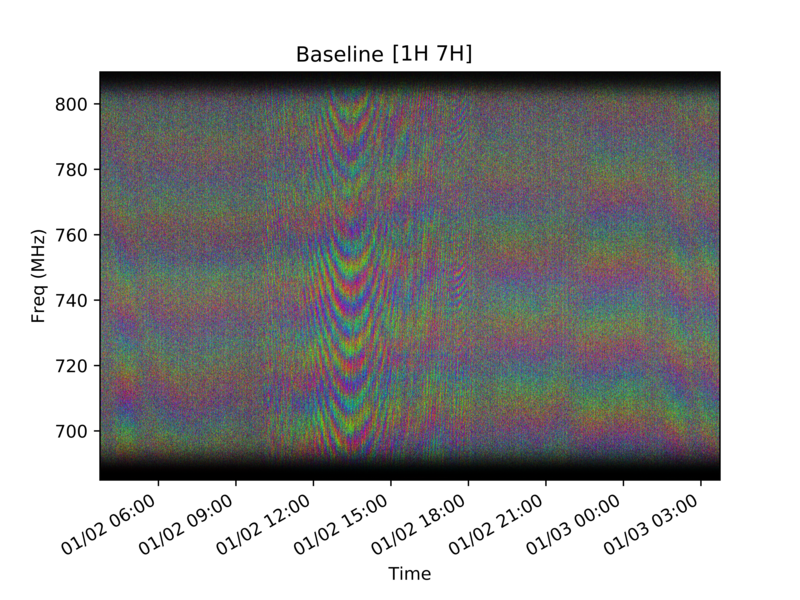}
    \caption{The complex visibility after smoothing the components of the largest eigenvector in spherical coordinates (described in Sec. \ref{sec:smoothing}, before scaling by a gain factor, %$\chi^2$ optimization 
    described in Sec. \ref{sec:scaling} )
    for four different typical baselines. Nightly mean subtraction has been applied. There is still some signal from the Sun that has not been removed.
    }
    \label{fig:aftersmoothing}
\end{figure*}

%\section{Subtracting the Sun signal}
\section{Improving the Sun subtraction}

%\rzorange{\st{ I suggest the following title for this section: | Subtracting the sun signal}}

\label{sec:leftover}
As we can see from Fig.~\ref{fig:zero_eigseparation}, some of the solar signal still remains in the visibility matrix, mainly the signal that leaked to the second largest or even to the third largest eigenvalue. We attempt to remove this residual signal through the following 2 steps.  

\subsection{Smoothing the eigenvalues and eigenvectors}
\label{sec:smoothing}

%\rzorange{\st{I think that the smoothing procedure tries to address the impact of instrumental noise, not the fact that multiple sources in the sky contribute and affect thus the largest eigenvalue-eigenvector. This is what it is said in the second paragraph. The first sentence here might thus be misleading. } }{\color{blue} I modified the text slightly. However I am not exactly sure if everything is addressed properly. May be you can check and rewrite whatever is necessery}

The problem with the direct eigenvalue removal method described above is that it is based on the assumption that there is a single  source on the sky. 
This is not true in this case, as the visibility matrix contains signals from other sources as well as instrument noise. In Fig.~\ref{fig:eigen_vector_smoothing},  one component of the eigenvector corresponding to the largest eigenvalue is shown in light green over a period of 4 hours. The data, sampled every second, are noisy. However, as the Sun  moves smoothly and the beam is not expected to be structured on small scales, we expect the eigenvector to vary smoothly with time. 
These fluctuations in the eigenvalue probably come from noise.
The long term (minute level and longer) fluctuations originate in the structure of the sidelobes of the telescopes. 

As the noise in the visibility matrix may cause the Sun signal to leak from the largest eigenvalue to other eigenvalues, in this section we try to reduce the effect of noise. For doing that, we fit a smooth curve (black line) through the  eigenvectors corresponding to the Sun signal. 
%\footnote{{\color{red}Please describe the smooting algorithm in brief. Did you take a fourier transform and remove he high frequency parts? or Are you taking an average of some data points. Please discuss briefly}}. 
This smoothed signal from the largest eigenvector is then subtracted from the original visibility to construct the Sun-removed visibility. 

The cleaning routine can be summarized as follows. The visibility matrix $\mathbf{V}^{(X)}$
%, whose element $V^{(X)}_{ij\,(t,\nu)}$ represents the visibility between feed $i$ and $j$, we calculate 
is first decomposed into the eigenvalue and the eigenvectors for each time and each frequency bin,

\begin{equation}
    \mathbf{V}^{(X)} = \mathbfcal{E}^{(X)} \mathbf{\Lambda}^{(X)} \mathbfcal{E}^{(X)~-1},\qquad X=\{H,V\}.
\end{equation}

\noindent Suppose $\mathcal{E}_{S}$ is the eigenvector corresponding to the largest eigenvalue, $\lambda_S$. As shown in Fig.~\ref{fig:eigen_vector_smoothing}, the direction of $\mathcal{E}_{S}$ will vary in every second. 
%\rzorange{Is this an edit leftover ? [As we have already choosen]} 
The $n$-dimensional complex eigenvectors have only $2n-1$ degrees of freedom as we have already set the first component to be real and positive. 
As the eigenvectors are unit vectors, the total number of independent components becomes $2n-2$. If we fit a smooth line through each of the $2n-1$ components, then we will overfit and the amplitude of the eigenvectors will not be 1. To keep the eigenvector normalized while doing the fitting, we express each (complex) component of the eigenvector in $n$-dimensional spherical coordinates and %$\mathcal{E}_{S}$, corresponding to the largest eigenvalue, $\lambda_S$, in polar coordinates. 
%The subscript $S$ denotes the Sun, which is the dominant contributor to the largest eigenvector. 
then  fit a smooth line through the tangents of the angles in spherical coordinates and convert back to Cartesian space. This gives the black line, shown in Fig.~\ref{fig:eigen_vector_smoothing}. Smoothing in spherical coordinates ensures that the normalization of the eigenvector is preserved during the smoothing procedure. (See Appendix.~\ref{App:1} for details.)

% \begin{equation}
%     x_i = \prod_{j=1}^{i-2} \sin(\theta_j)\cos(\theta_{i-1}) \hspace{1 cm}\forall i \in [1,2n-2]
% \end{equation}

%We can reasonably assume that the Sun signal and the array's beam vary smoothly as the Earth rotates, and the small second level fluctuations are due to the external noise, RFI, etc. If we fit a smooth curve through the eigenvector, then we can remove some of the contribution from the noise. 

Let the smoothed components of the largest eigenvectors be $\tilde{\mathcal{E}}_{S}$. If $\tilde{\lambda}_{S}$ is the contribution to the visibility from the direction of the eigenvector $\tilde{\mathcal{E}}_{S}$, then we can write, $\tilde{\lambda}_{S} =\tilde{\mathcal{E}}^{T}_{S}\mathbf{V}\tilde{\mathcal{E}}_{S}$. If we assume that this smoothed component comes from the Sun signal, then the contribution to the visibility from the Sun is given by 
\begin{equation}
\tilde{\mathbf{V}}_{S}=\tilde{\lambda}_{S}\left[\tilde{E}_{S} \otimes \tilde{E}_{S}\right].
\end{equation}
After subtracting the Sun signal, the contribution to the visibility from the rest of the radio sky and noise is given by 
\begin{equation}
\label{rm_Sun}
\mathbf{V}_{\text{Sky}}=\mathbf{V}-\tilde{\mathbf{V}}_{S}\,.
%=\mathcal{E} \mathbf{\Lambda}\mathcal{E}^{-1}-\mathbf{V}_{S}
\end{equation}
In Fig. \ref{fig:aftersmoothing} we show the complex visibility after removing the Sun signal using this particular algorithm. In comparison to the simplest algorithm, of just removing the largest eigenvalue, this new algorithm works better. However, we can see that some of the Sun signal is still present in the visibility. 

\begin{figure*}
    \centering
    \includegraphics[width=0.9\textwidth]{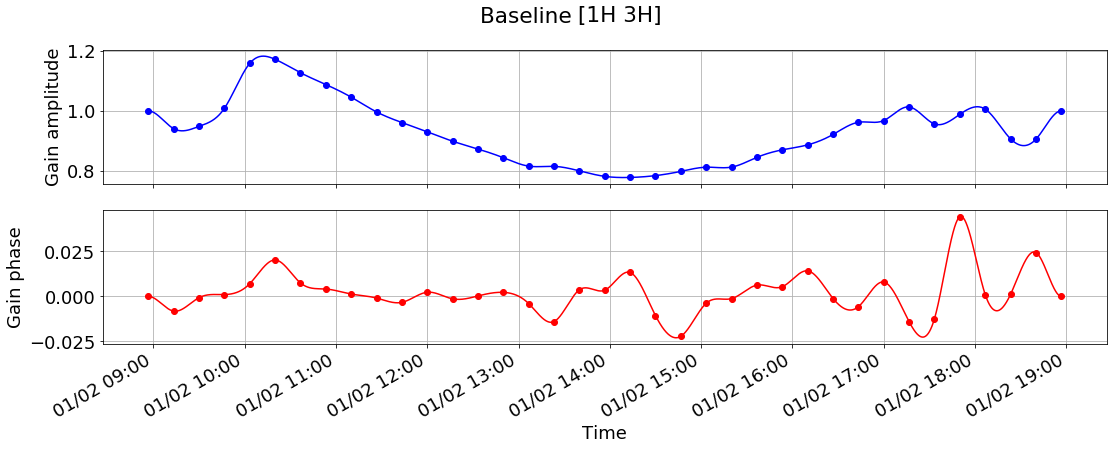}
    \caption{Plot of the gain $g = A e^{i\phi}$ after $\chi^2$ minimization for 10 hours during the daytime. Blue: Plot of the gain amplitude $A$. Red: Plot of the gain phase $\phi$ in radians. The dots represent the points of $\chi^2$ minimization that occur every 1000 seconds. These gain values are extrapolated to the intervening points for a total of 36,000 seconds.}
    \label{fig:chisquare_gain}
\end{figure*}

\subsection{Scaling the Sun signal from eigenvalue analysis}
\label{sec:scaling}

The above eigenvalue analysis is based on the assumption that the signal coming from Sun is contained in the largest eigenvalue. As discussed before, this assumption is not correct because of the leakage of power into other eigenvalues. %There are many reasons that may cause leakage of Sun signal to other eigenvectors. 
%Firstly, the noise and the external signal in the visibility matrix may lead to the leckage of power from the largest eigenvalue to the other eigenvalues as we have discussed in the previous section. 
%Apart from that, in the eigenvalue analysis we don't have the exact value of the auto correlation signal, and we discussed that the autocorrelation signals are replaced by a proxy value shown in Eq.~\ref{autocorr}, for  the analysis. This may cause some leakage of the solar signal to other eigen modes. Therefore, we need to account for that effect.
%Some of the solar signal may also get reflected on the ground and then gets detected through the far sidelobes. 

%In addition, there are other sky signals and noise present in the data. Therefore, the eigenvector corresponding to the largest eigenvalue may not orient exactly in the direction of the Sun. {\color{red}There can also be slight physical misalignment in the dish configuration.} Therefore, we may not get the full contribution form the Sun in largest eigenvector. 

%In Fig.~\ref{fig:eigen_vector_secondlargest}, we have shown the phase from one of the components of the second largest eigenvector. We can see the fringes during daytime, ensuring that some of the solar signal is leaked to the second largest eigenvector or a even smaller amount to third largest eigenvector. 

To overcome this issue, we consider that during the daytime the signals from the sky are much smaller than the solar signal. Therefore, the Sun signal, $\mathbf{V}_S(t,\nu)$, calculated from our analysis should roughly match with the visibility $\mathbf{V}(t,\nu)$ during the daytime as the other signals are negligible in comparison to the $\mathbf{V}_S(t,\nu)$, provided that there are no other strong sources during the day. To do that we introduce a scaling (gain) factor, $g = Ae^{i\phi}$, for each 1000 seconds (about 15 min) of daytime data and minimize 

\begin{eqnarray}
\chi^2 &=&\sum_{t,\nu}\left[\Re(\mathbf{V}-g\mathbf{V}_{S}(t,\nu))\right]^2 \nonumber\\
&+&\sum_{t,\nu}\left[\Im(\mathbf{V}-g\mathbf{V}_{S}(t,\nu))\right]^2.
\end{eqnarray}

\begin{figure*}
    \centering
    \includegraphics[width=0.49\textwidth,trim = 0 .1 1 1, clip]{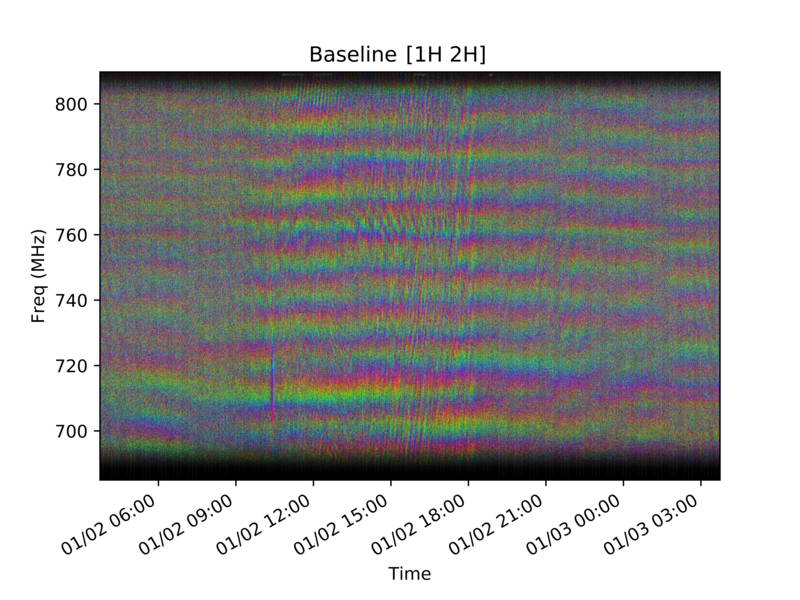}
    \includegraphics[width=0.49\textwidth,trim = 0 .1 1 1, clip]{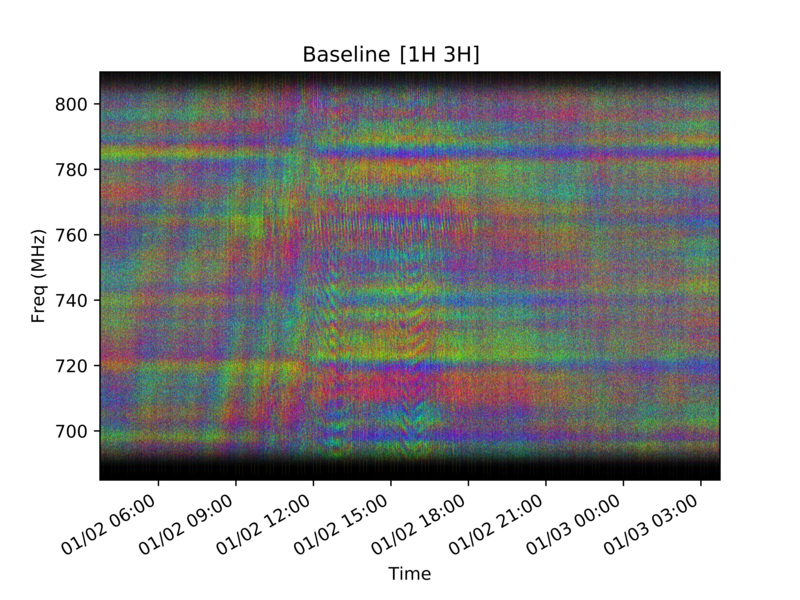}
    \includegraphics[width=0.49\textwidth,trim = 0 .1 1 1, clip]{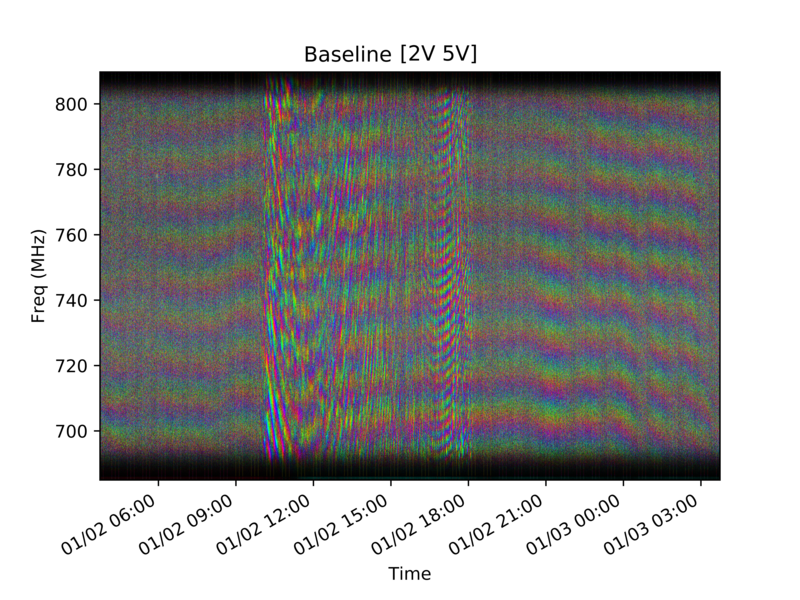}
    \includegraphics[width=0.49\textwidth,trim = 0 .1 1 1, clip]{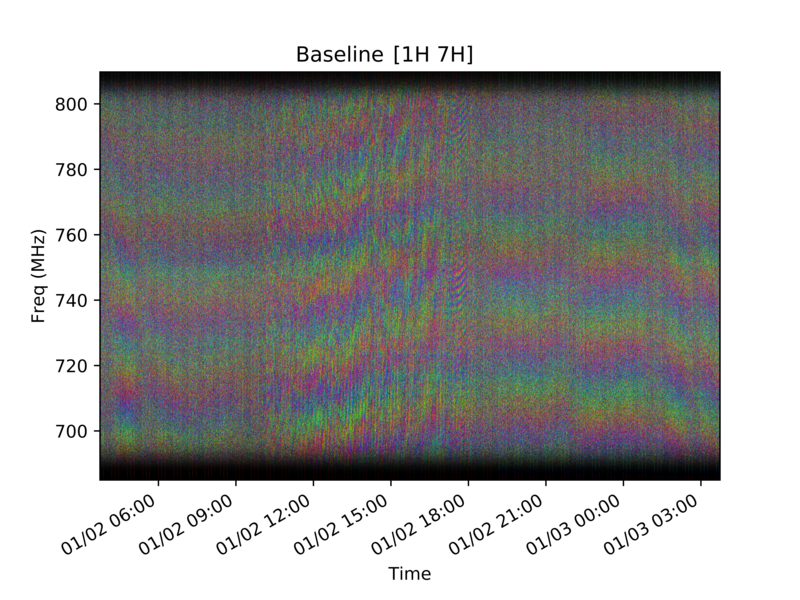}
    \caption{
    The waterfall plot of the complex visibility after $\chi^2$-optimization 
    scaling the Sun signal by a gain factor (Section \ref{sec:scaling}) for four typical baselines. %\rzorange{I suggest to replace  the next two sentences of the caption by:  } 
    We can see that most of the sun signal contribution has been removed. However, very weak noisy fringes are still visible during daytime, specially in some baselines, such as in [2V 5V], shown in the bottom left panel.
    % We can see that in most of the baselines most of the sun signal is getting completely removed. However, for some of the baselines, we can still see very weak noisy fringes during daytime. 
    }
    \label{fig:finalValue}
\end{figure*}

{\noindent Here $\Re(\,)$ and $\Im(\,)$ are the real and imaginary parts of the quantity inside the bracket. We get 36 gain factors ($g$), calculated from $10$ hours ($36,000$ seconds) of daytime data.
In Fig.~\ref{fig:chisquare_gain}, we show the plot of $g$ over 10 hours of daytime, with circular dots. The smooth lines show the interpolated data. We can see that $g$ varies smoothly throughout the day. The expectation is that the $|g|$ should be very close to $1$ and very smooth, and the phase variation should be very small. This is because the Sun and the sky move smoothly through the beams over the day. As long as the Sun signal is strong enough in comparison to the background sky we can expect that the power leakage will vary smoothly and the gain variation should also be smooth. As the signal in the largest eigenvector and leaked power both are coming from Sun we can expect the phase variation to be minimun. Fig.~\ref{fig:chisquare_gain} shows that the assumption is a good one in this case. However, near sunrise and  sunset the amplitude and the phase change rapidly, possibly because the Sun signal is weaker at those times. }

The interpolated $g$ is used as a multiplication factor to %$\tilde{\mathbf{V}}_S(t,\nu)$ and treated as the
determine the solar contribution $g\times\tilde{\mathbf{V}}_S(t,\nu)$, which is finally subtracted from $\mathbf{V}(t,\nu)$. This gives our final Sun-removed signal from the daytime data, i.e. 

\begin{equation}
\mathbf{V}_{\text{Sky}} (t,\nu)=\mathbf{V}-g_{\text{int}}(t,\nu)\tilde{\mathbf{V}}_{S}(t,\nu)\,.
%=\mathcal{E} \mathbf{\Lambda}\mathcal{E}^{-1}-\mathbf{V}_{S}
\label{Eq.FinalSunRemoval}
\end{equation}

In Fig.~\ref{fig:finalValue} we show the complex visibility after the solar contamination removal using Eq.~\ref{Eq.FinalSunRemoval}.  We can see by visual inspection that most of the contamination signal is removed and the fringes from the weaker sources in the background sky are visible. However, for some of the baselines, a significant fraction of the Sun in the form of weak fringes still remains, as we can see in the bottom left plot. In the next section,
%we will quantify how much Sun signal has been removed.
we will make a first estimate of the performance of our solar signal subtraction and its effect on the signals from the fainter sources.

\section{Understanding the efficiency of the algorithm}
To test the efficiency of the above algorithm (AlgoSCR), we apply it to simulated data sets. This allows us to check the fraction of the solar contaminant signal that is removed and how much of the sky signal we are erroneously removing by the analysis. 

\subsection{Construction of simulated data}
For constructing a simulated visibility signal $\mathbf{V}_\text{sim}$, we assume that the electric field at each feed antenna contains contributions from Sun, the sky, and noise. %Its not completely irrelevant to assume that the noise depends on the signal from external source, i.e. the noise is high during daytime and small during night. However, Tianlai data don't show any such feature. Therefore during our analysis 2
We have assumed that the noise variance is the same throughout the analysis. 

\begin{figure*}
    \centering
    \includegraphics[width=1.0
    \textwidth,trim = 0 35 1 20, clip]{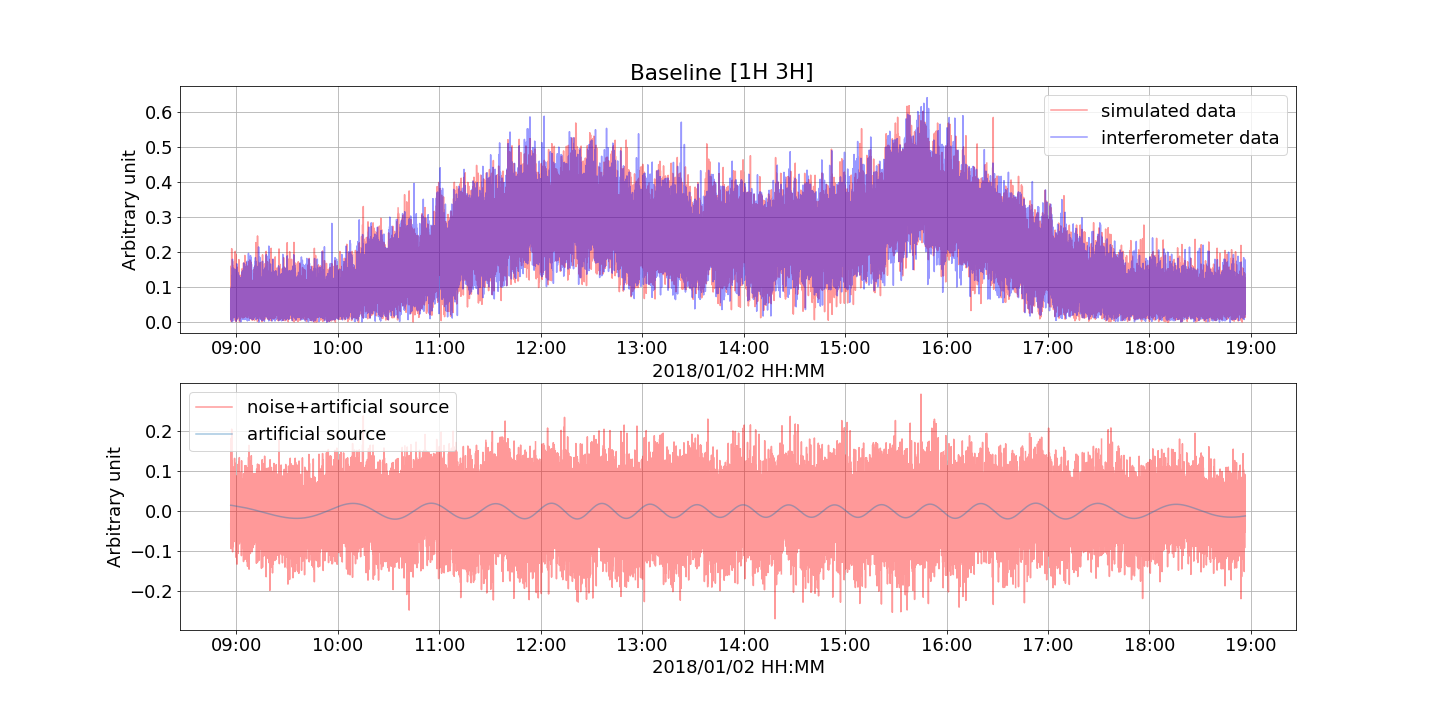}
    \caption{Top: The amplitude of the visibility for baseline [1H 3H]. The actual data from the Tianlai dish array is colored blue, while the simulated data are shown in red. The actual data are very similar to the simulation;  regions of overlap appear purple. Bottom: In red, the real part of the noise plus the artificial sources. The blue line shows the real part of the signal from the artificial sources that is added to the data. The imaginary part is similar to the real part and is not shown. %{\color{red} Should we remove the opacity?}
    }
    \label{fig:sim_data}
\end{figure*}

The receiver noise is modeled as Gaussian noise in the electric field, $E_{\text{noise}\,i}$, at the feed antenna.  We consider the noise contribution to the electric field to be Gaussian in each sample.  
%\rzorange{I think we should be more precise and quantitative here. We should have a noise level $T_{sys} \sim 100 K$ and the gaussian random noise $\sigma$ added in 10 ms interval should correspond to $ \sim T_{sys}/\sqrt{\delta \nu \delta t}$. 
%We need to make sure that the sun signal has the correct amplitude with respect to $T_{sys}$. It is said that the sources are 10 times smaller than the noise. But that would be huge 10 K contribution from the sky sources ? I guess, you mean that the sources are 10 times smaller than the noise fluctuations, then you have to specify the frequency bandwidth and integration time. Looking at the plots, assuming 1 s. integration time and 1 MHz bandwidth, I see noise fluctuations around 0.1 (which should then correspond to 100 mK, and sun signal around 0.3 ( 300 mK ), and then source contribution around 10 mK ?  Can you check and include numbers in the text ?  }

In the Tianlai dish array the integration time in the correlator is $1$-sec. The correlator takes in the data that are collected every few microseconds and averages them in an interval of $1$-sec. To simulate this process, we add Gaussian random noise in the electric field with a sampling interval of 10~ms. %In the integration time of $\tau_{\texttt{int}}=1$ second, 
We then calculate the noise contribution to the visibility as  
$ V_{\text{noise}\,(i,j)}\equiv\langle E_{\text{noise}\,i}^* \,E_{\text{noise}\,j} \rangle_{\tau_{\texttt{int}}}$, where  $\langle\;\;\;\rangle_{\tau_{\texttt{int}}}$ represents the ensemble average over integration period, $\tau_{\texttt{int}}=1$ second.  Here we have $100$ data points for every second on which the average is carried out. This method also ensures that the autocorrelation visibilities follow a $\chi^2$ distribution and the cross-correlation visibilities follow a product normal distribution. %Therefore, the sky signal is also included in the noise. We assume the noise in the electric field is Gaussian. 
The mean and variance of $E_{\text{noise}\,i}$ are chosen empirically so that the simulated visibility, $\mathbf{V}_\text{sim}$, matches the observed visibility. The mathematical details on how to calculate the visibilities from artificial point sources in the sky are shown in Appendix \ref{App:2}.
%\begin{equation}\label{noise_vsb}
%    V_{\text{noise}\,(i,j)}\equiv\langle E_{\text{noise}\,i}^* \,E_{\text{noise}\,j} \rangle_{\tau_{\texttt{int}}}
%\end{equation}
 
\begin{figure*}
    \centering
    \includegraphics[width=1.0
    \textwidth,trim = 0 100 1 100, clip]{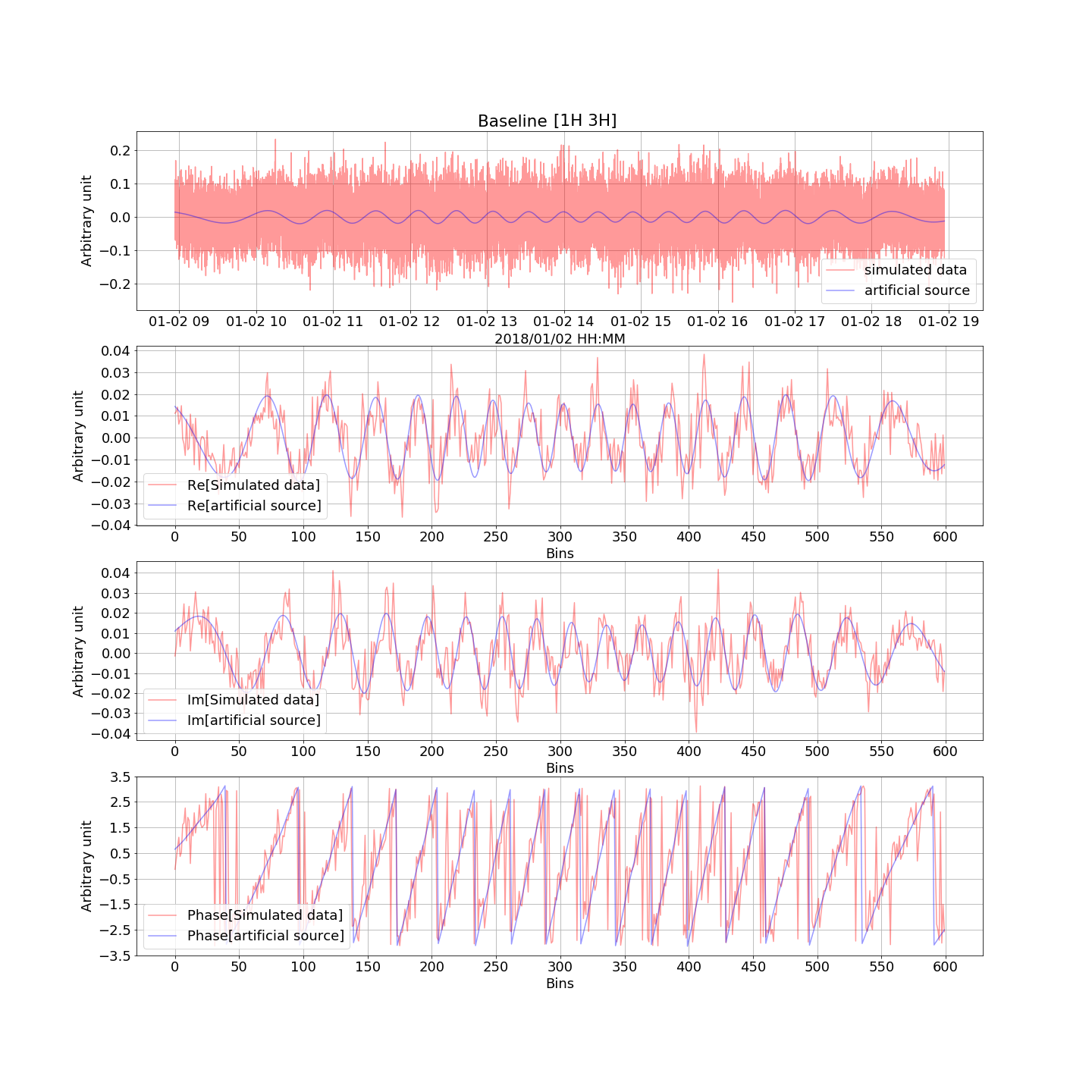}
    \caption{\label{fig:sim_Sun_removed} Top: the real part of the Sun-removed visibility from simulated data for baseline [1H 3H].  The signal from the artificial sources is shown in blue. The visibilities are sampled every second. The lower three plots show the real part, imaginary part, and phase of the Sun-removed visibility, after 60-second averaging.
    }
    
\end{figure*}
\begin{figure*}
    \centering
    \includegraphics[width=1.0
    \textwidth]{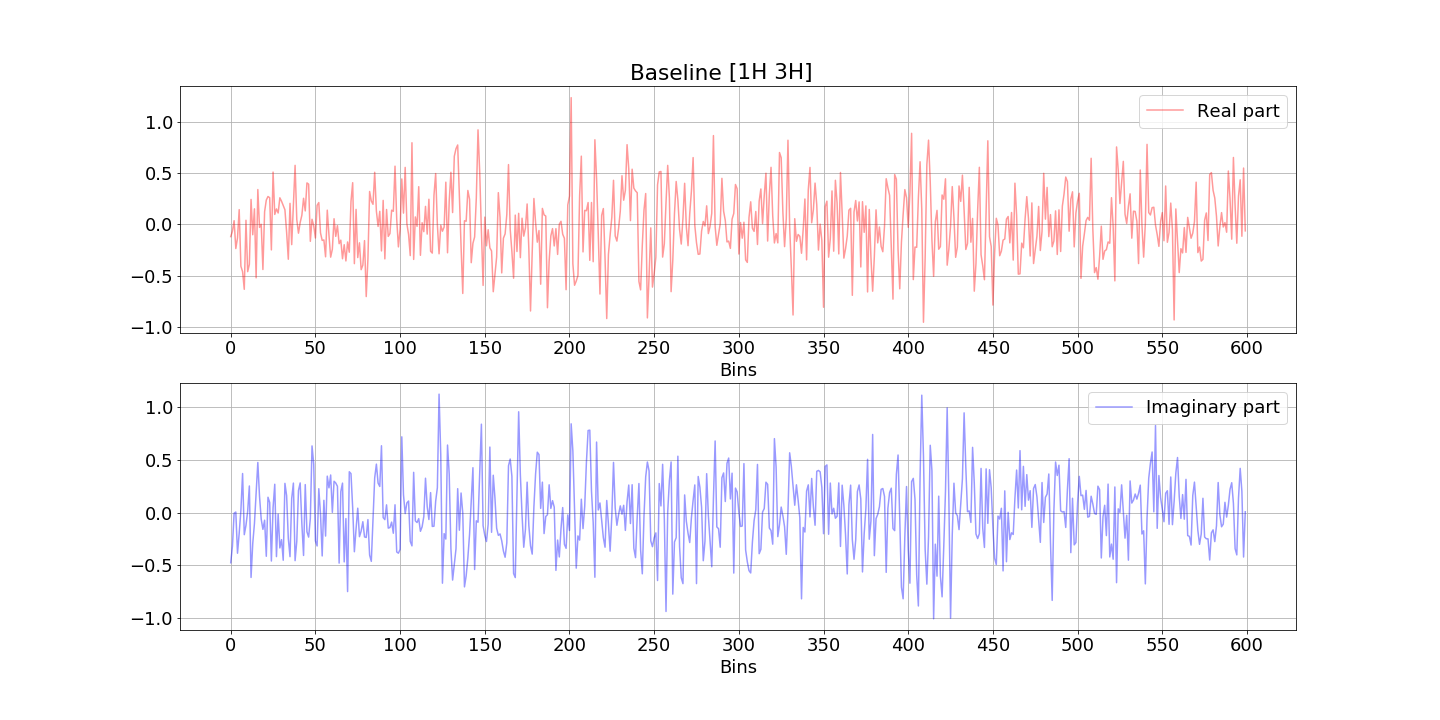}
    \caption{\label{fig:sim_Sun_removed_frac1}  Plot of the real and imaginary part of  $(V_\text{sim}-V_\text{org})/\sigma$, where $\sigma^2$ 
    %\sim 2 T_\text{sys}/\sqrt{\Delta\nu\Delta t}$ 
    is the variance of the added noise. We can see that the ratios for both the real and imaginary parts are roughly within $3$. As the noise is Gaussian, we can expect that the noise signal should be with $3\sigma$. we conclude that the Sun removal algorithm does not introduce any significant additional noise in this analysis.% We assume that the system temperature is $T_\text{sys} = 100$ K.}
    %{\color{blue} Need to Check. There may be something wrong.} {\color{green} Changed the system temperature to the variance of the noise.}{\color{blue}What is the system temperature now???}{\color{green} $\sigma^2(1s)$=0.0039, $\sigma^2(1min)$=6.61\times10^{-5}}
    }
    
\end{figure*}

% {\color{red} Can you please give the mathematical details of the three made up sources. I mean what are there locations, how are you calculating their visibilities etc. }

 To create the simulated Sun signal, we have taken the largest eigenvalue and correponding eigenvector from Eq. \ref{vis_Sun} from the Tianlai dish array data and treated it as the solar signal. The electric field for the Sun, thus calculated, is added to the simulated noise.

%\begin{equation}
%    \mathsf{E}_S(\nu,t) = \sqrt{\lambda_S} E_S
%\end{equation}

%The value of $\mathsf{E}_S^{(H)}$ at each second is then extrapolated to 10 ms, so in each second there will be 100 $\mathsf{E}_S^{(H)}$ extrapolated data point. The random Gaussian noise is then added to the electric field each 10 ms. We also added artificial sources near the north pole and calculated the theoretical visibility generated by those sources. The total simulated electric field at each dipole is given by
%\begin{equation}
%    \mathsf{E}_\text{sim} = \mathsf{E}_S + \mathsf{E}_\text{noise}+ \mathsf{E}_\text{artificial source}
%\end{equation}

For the simulated artificial sources,  we assume the telescope array is pointed at the NCP. The artificial sources are three made-up sources near the NCP.  All the artificial sources are visible within the main beam, which is assumed to be Gaussian.  Their brightnesses are chosen so that the amplitude of their combined visibility is about 10 times smaller than the noise. (An analysis with different source strengths is presented in the next section.) The artificial source visibilities are frequency- and baseline-dependent, just as visibilities from real sources on the sky. We also assume that the visibilities for the Sun and artificial sources  are uncorrelated, i.e., there is no cross-term between the Sun and the artificial sources. This makes the visibilities for the Sun and artificial sources additive, as shown in Eq. \ref{Vsim}. Please check Appendix~\ref{App:2} for details.

{\noindent We can also assume that the visibilities for the Sun and artificial sources  are uncorrelated, i.e., there is no cross-term between the Sun and the artificial sources. This makes the visibilities for the Sun and artificial sources additive, as shown in Eq. \ref{Vsim}. Please check App.~\ref{App:2} for details. }

\begin{equation}\label{Vsim}
    \mathbf{V}_\text{sim} =  \mathbf{V}_\text{noise} +
     \mathbf{V}_\text{S} + \mathbf{V}_\text{artificial sources} 
\end{equation}

\begin{figure*}
    \centering
    \includegraphics[width=1.0
    \textwidth]{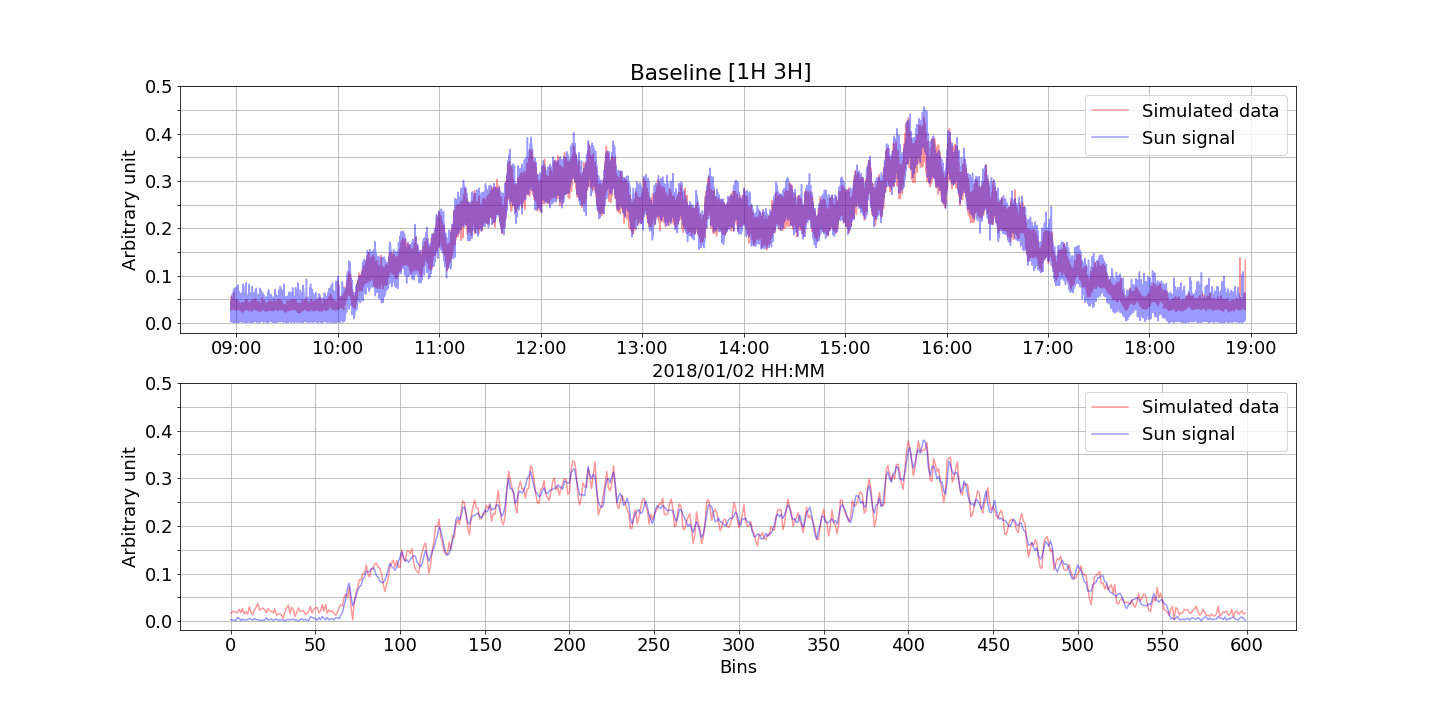}
    \caption{Top: The difference between the simulated visibility (shown in Fig.~\ref{fig:sim_data}) and Sun-removed visibility (shown in Fig.~\ref{fig:sim_Sun_removed}) is shown in red. The signal from the largest eigenvector, which is used as the Sun signal during the daytime is shown in blue. Bottom: Same plot over the same time interval but the data are averaged in 60 second time bins to reduce the noise. }
    % {\color{blue} Need to fix the labels.} {\color{green}label1: Recovered Sun signal, label2: Input Sun signal}}
    \label{fig:sim_diff}
%\end{figure}
%\begin{figure}

%     \centering
%     \includegraphics[width=1.0
%     \textwidth]{figs/frac_diff.png}
%     \caption{Top: the fractional difference between the simulated Sun signal (shown in red on top of Figure \ref{fig:sim_diff}) and the signal from the largest eigenvector, which is assumed to be the Sun signal during the daytime. Bottom: Same plot over the same time interval but the data are binned in 60 second time bins.{\color{blue} Do we really need these plots? Its confusing as the percentage difference will not tell anything, its a random number. Also, its not adding anything to the paper, as we already prove the point in the previous plot.}}
%     \label{fig:frac_diff}
\end{figure*}

\subsection{Results from the simulated data}

We generated simulated data as shown in Fig.~\ref{fig:sim_data}. The top panel of Fig.~\ref{fig:sim_data} shows the amplitude of the visibility for baseline [1H 3H] for simulated and Tianlai dish array data: The plot in blue shows the actual complex visibilities from the Tianlai dish array, and the red plot shows the simulated data in our simulation (see Equation \ref{Vsim}).  The bottom panel shows the real part of the signal from the artificial sources (in blue). 
%amplitude of the 
The real part of the combined signal (simulated noise and the visibility of the artificial sources that is added to the Sun) is shown in red. 
%. %The simulated noise (in red) is % consists of random Gaussian noise  %.

%The simulated data matched with the interferometer data to a few percent.

\begin{figure*}
    \centering
    \includegraphics[width=0.45\textwidth,trim = 1 10 1 10, clip]{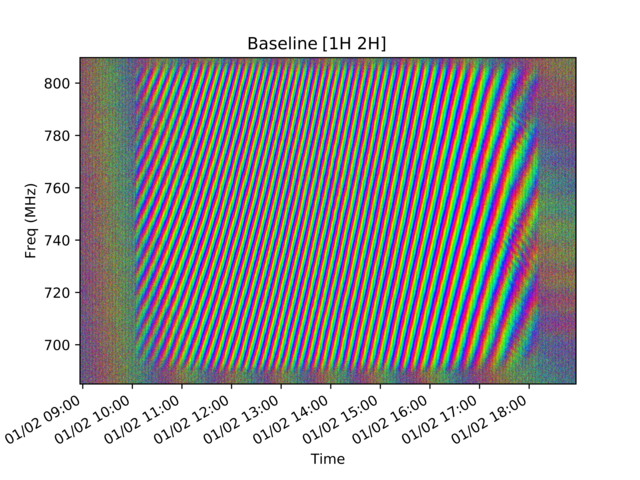}
    \includegraphics[width=0.45\textwidth,trim = 1 10 1 10, clip]{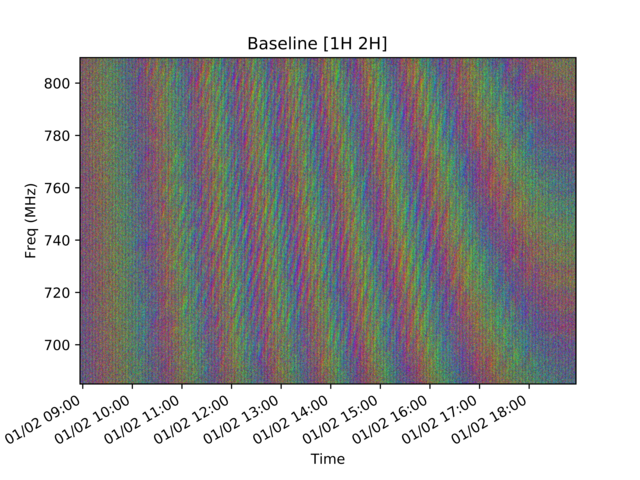}
    \includegraphics[width=0.45\textwidth,trim = 1 10 1 10, clip]{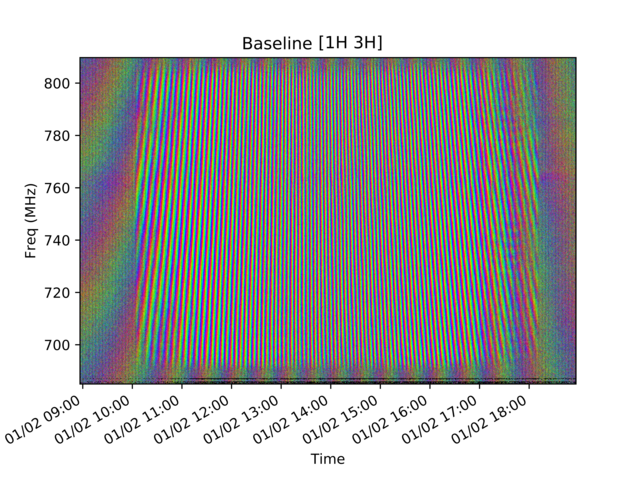}
    \includegraphics[width=0.45\textwidth,trim = 1 10 1 10, clip]{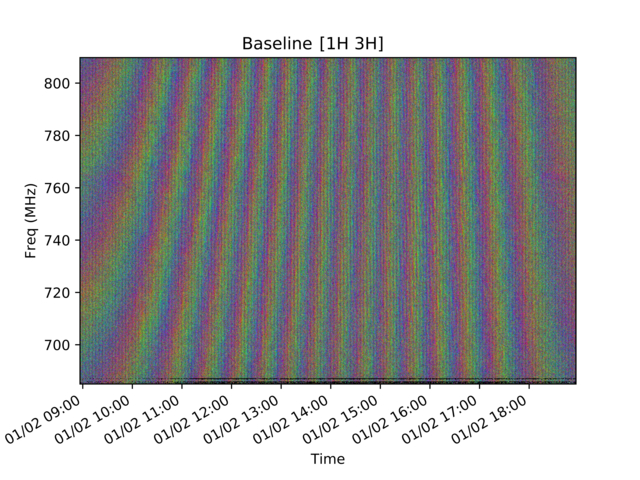}
    \includegraphics[width=0.45\textwidth,trim = 1 10 1 10, clip]{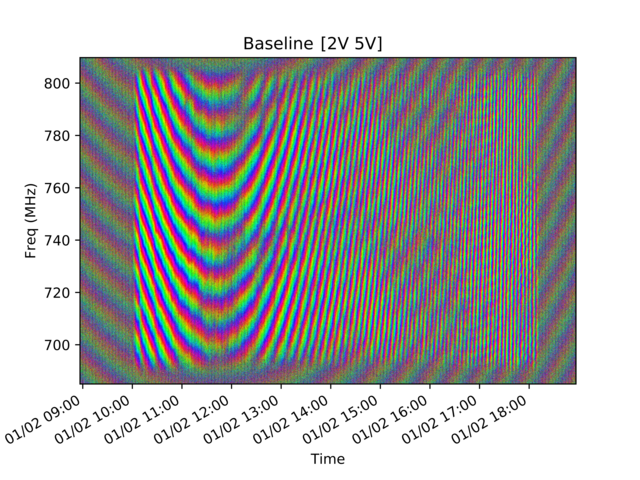}
    \includegraphics[width=0.45\textwidth,trim = 1 10 1 10, clip]{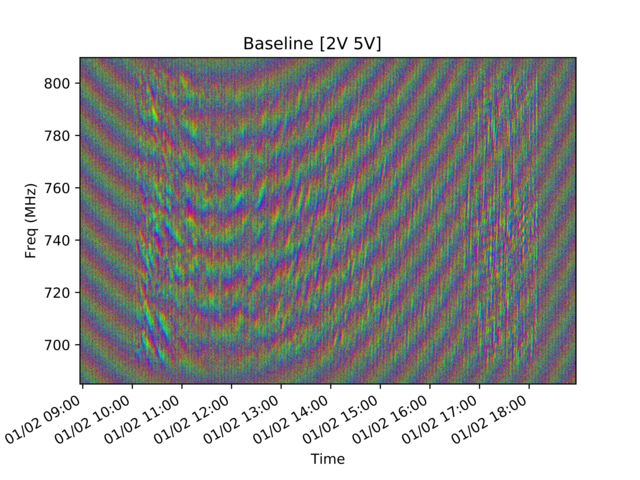}
    \includegraphics[width=0.45\textwidth,trim = 1 10 1 10, clip]{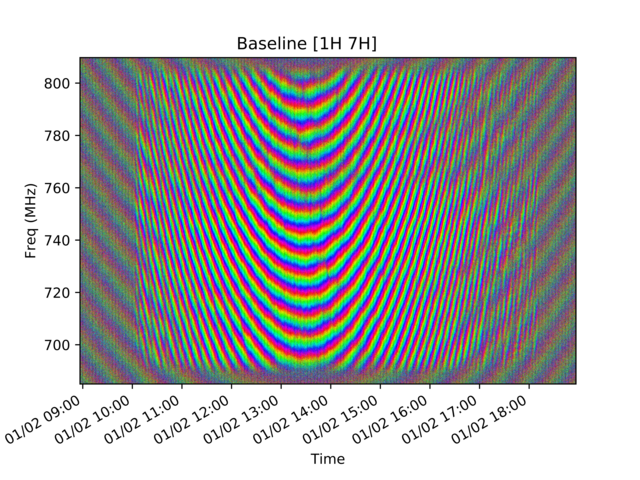}
    \includegraphics[width=0.45\textwidth,trim = 1 10 1 10, clip]{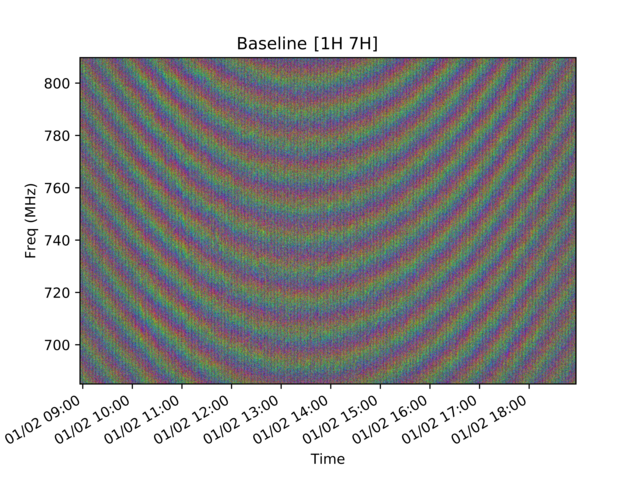}
    \caption{
    Left column: Simulated daytime visibilities for several baselines. The bright fringes are from the Sun and the weaker fringes are from the artificial sources.  Right column: Sun-removed visibility using AlgoSCR. The fringes from the artificial sources are clearly visible. We can see that most of the solar contamination is removed by AlgoSCR.
    }
    \label{fig:sim_removal}
\end{figure*}

 %The data after the Sun signal removal is shown in the 
We apply AlgoSCR to the simulated data.  The top panel in Fig.~\ref{fig:sim_Sun_removed} shows the real part of the visibility (in red) after applying the Sun removal algorithm, along with the real part of the visibility of the artificial sources (in blue) for baseline 40 [1H 3H]. As the nature of the imaginary part will be similar, we have not explicitly shown it in the plot. In the bottom three panels, the visibilities of the Sun-removed simulated data and artificial sources are binned in 60 second bins. The 60 second binning is used to increase the signal-to-noise ratio. We see that the real and imaginary parts of the Sun-removed signal closely resemble those from the artificial sources and that the phase is not affected.
 
%  {\color{red}all should be plotted in 60 sec bin. Which figures should I change?}
 
%{ \color{red} 
The ratio of the difference between $V_\text{sim}-V_\text{org}$, and the noise standard deviation, $\sigma$, is plotted in Fig.~\ref{fig:sim_Sun_removed_frac1}. We can see that the ratio is within 3. As the injected noise is Gaussian,  we can expect that most of the visibility should also fall within $3\sigma$. Therefore, Fig.~\ref{fig:sim_Sun_removed_frac1} ensures that the recovery of the signal using AlgoSCR does not introduce additional noise.

%We see that the difference on average is about half the expected fluctuation from noise. }
%Here we should note that we can relate the $\sigma$ with the system temperature ($ T_\text{sys}$) as  $\sigma = 2 T_\text{sys}/\sqrt{\Delta\nu \Delta t}$, provided we convert the visibilities in temperature unit by multiplying them with the antenna gain. However, the plots that we have shown here are not multiplied with the antenna gain and are in arbitrary units. To match the simulated visibilities with the real data we choose $\sigma = 0.06$ 
%We used the 60~s time bins ($\Delta t = 60$s), single frequency bins ($\Delta \nu = 244$ kHz), and assume that the system temperature $T_\text{sys} = 100$~K. 

%removed visibility, i.e.{\color{red} $\frac{V_\text{sim}-V_\text{org}}{V_\text{org}}$ }. The difference is within 10 percent, showing that the source that was introduced in the analysis did not get removed by our Sun removal algorithm. The spikes are coming because the $V_{org}$ is $0$ at those places.

%  {\color{red} is it just the difference or the difference/signal *100\%  ---- just the fractional difference (no percentage), you can see this in Figure \ref{fig:sim_Sun_removed}}

% {\color{blue} Probably its not a good representation of the data. I think we should remove Fig~22 and Fig~24. Or may be we can discuss with the group if some has some new and better idea of representing the data. !!! }

% shows two different ways to estimate the signal from the Sun.  
The red plot in Fig.~\ref{fig:sim_diff}, shows the difference between the amplitude of the simulated visibility (including the Sun), and the visibility that we are getting after applying AlgoSCR. So this give the contribution from Sun in our simulated data. The blue curve is showing the Sun signal that we introduced for generating the simulated data. We can see that the plots match very well. Top plot is constructed using the data from each second and the bottom plot is after averaging the data over a minute.

%{\color{red} Figure \ref{fig:frac_diff} shows the fractional difference between the Sun-removed simulated data and the Sun signal. These plots show that the solar signal that was introduced in the data is completely removed by AlgoSCR. }

% {\color{red} is the again the percentage error or just the fraction ----- again it is just the fraction }

%Therefore, this full analysis shows the effectiveness of the algorithm. 

In the next set of plots, Fig.~\ref{fig:sim_removal}, we show the complex visibilities for some of the baselines, before and after the solar contamination removal by AlgoSCR. The visibility data  show that the artificial sources that we had introduced are clearly visible after the solar signal removal, even though the source strength was much smaller than the Sun signal and the noise. This simulation shows the potential of the Sun removal algorithm.

\subsection{Comparing efficiency of the method for different external source strengths}

%\rzorange{Minor remark : I think that this section would better fit as the last sub-section of the last section (5.3) \\
%The discussion is based on figure 26.  I would represent in a single figure with two panel, a reduced $\chi^2$ computed as a relative difference (you divide $|V_{org} - V_{sim}|^2$ by 
%$| V_{org}|^2 $) and a second reduced $\chi^2$, where you have normalised by the noise fluctuation $ | V_{noise} |^2$. 
%I think that you should look at this for time binning $ \gtrsim 60-100 s.$. We should see a noise dominated regime, at low source amplitude, then an intermediate regime where the source is small 
%(maybe 20 - 30 \% of the sun), when it is correctly recovered and not noise dominated, and then for bright sources, the fact that the bright source gets mixed with the sun.
%}

Here we compare the efficiency of AlgoSCR in recovering the artificial sources for different source strengths. For this analysis we use the real daytime visibility data taken by the Tianlai dish array as the base visibility. To this data we add the artificial visibility signal with different source strengths. We assume that sources are not correlated with the visibility data
%i.e. the cross-correlation between the Sun signal and the external source is $0$. Therefore,
and the visibilities are additive.

\begin{figure*}
    \centering
    \includegraphics[width=1.0
    \textwidth]{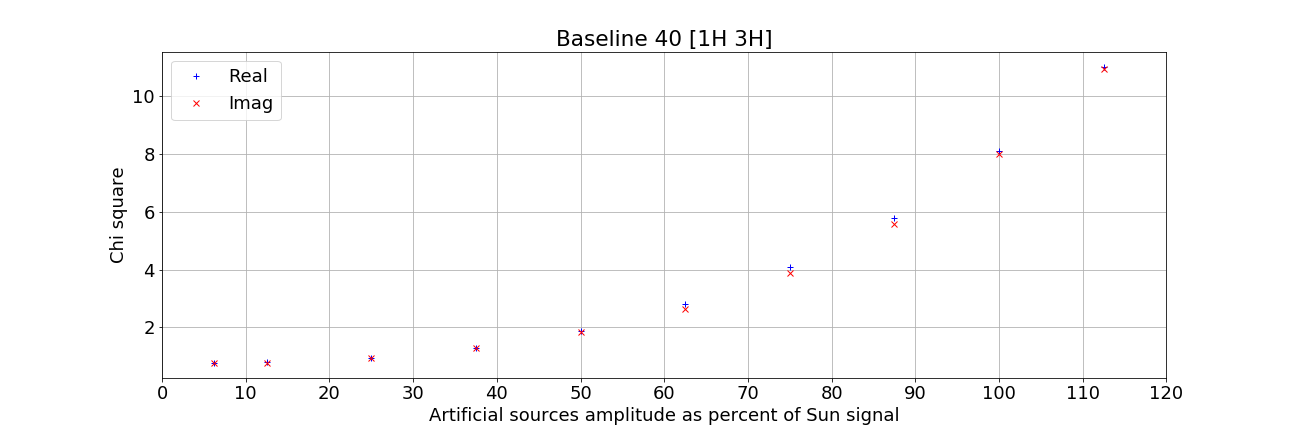}
    \caption{Plots of $\chi^2 = \frac{1}{n_t\sigma^2}\sum_{t,\nu}(V_\text{org} - V_\text{sim})^2$ from the real and imaginary parts of the visibility for different artificial source amplitudes. $n_t$ is the number of sample points in the time direction and $\sigma^2$ is the noise variance. The amplitude of the original visibility $V_\text{org}$ and the simulated visibility $V_\text{sim}$ are shown in Fig.~\ref{fig:sim_Sun_removed}. We can see that the $\chi^2$ is increasing as we increase the amplitude of the artificial source. %{\color{blue} Anh: I divide it by the number of seconds $\times$ variance $\sigma^2$. Santanu: Please don't remove the other plot. We need the $\chi^2$ plot. That is important. Both the plots are important for two different reason.}
    }
    \label{fig:chi_square2}
\end{figure*}

After running AlgoSCR to remove the Sun,  we using a $\chi^2$ statistic to compare the signal with the source visibility that was originally inserted.  The plot of the reduced $\chi^2$  for baseline [1H 3H] is shown in Fig.~\ref{fig:chi_square2} against the source strength.  Here $\chi^2$ is defined as $\chi^2 = \frac{1}{n_t \sigma^2}\sum_{t,\nu}|\mathbf{V}_{\text{org}}(t,\nu) - \mathbf{V}_{\text{sim}}(t,\nu)|^2$ during 10 hours of daytime. Here $\sigma^2$ is the noise variance and $n_t$ is the number of time-steps, which is the number of degrees of freedom in this case. As we are taking sampling each second for total of 10 hours, the number of degrees of freedom $n_t = 36000$. 

%{\color{blue} The noise variance, $\sigma$, is the same for all the points. Therefore, we ignored the $\sigma$ which just comes out of the summation and become an overall multiplicative constant.}

We can see the $\chi^2$ value is small for the cases in which the artificial source amplitudes are small compared to the Sun signal amplitude. As the artificial source strengths increase, the fit gets worse. This is because our analysis is based on the assumption that the solar signal is the only dominant signal. % The strengths of the other sources are very small compare to the Sun.
As the strength of the artificial sources increases, the assumption slowly breaks down. In such cases, the largest eigenvalue starts to capture signal from the artificial sources. When the artificial sources are larger than the Sun, the largest eigenvalue provides the contribution from the artificial sources and not the Sun. In such cases we are essentially removing the artificial sources and thus the $\chi^2$ grows quadratically.  

\begin{figure*}
    \centering
    \includegraphics[width=1.0
    \textwidth]{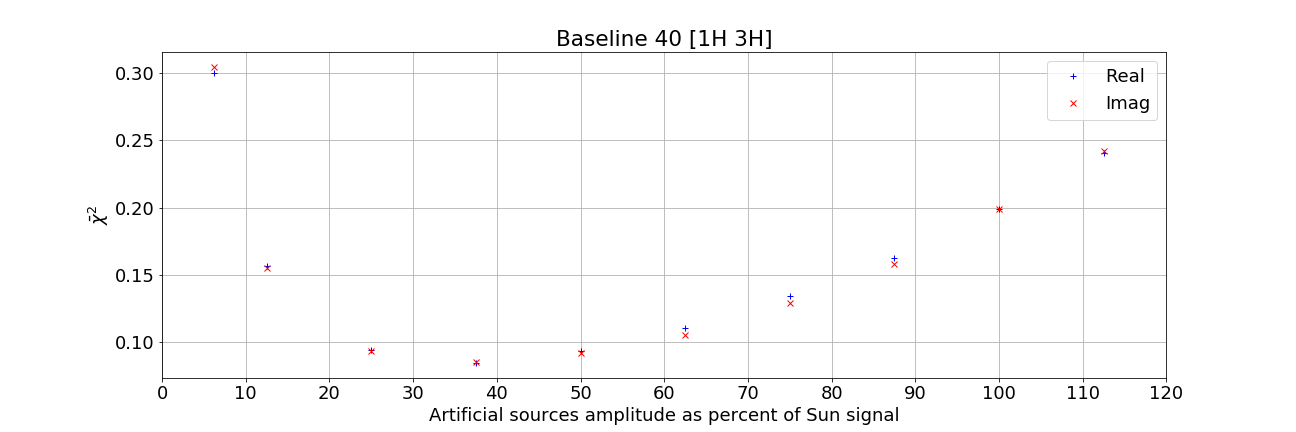}
    \caption{ Plots of $\bar{\chi}^2 = \frac{1}{n_t |V_{org}|}\sum_{t,\nu}(V_\text{org} - V_\text{sim})^2$ from the real and imaginary parts of the visibility for different artificial source amplitudes. Here, the plots are normalized by $|V_{org}|$ instead of $\sigma^2$. We can see that the $\bar{\chi}^2$ is lowest at about $40$, indicating that the recovery is best when the amplitude of the source is about $40\%$ of the Sun signal.}
    \label{fig:chi_square}
\end{figure*}

In Fig.~\ref{fig:chi_square} we have plotted the same $\chi^2$, 
where instead of dividing by $\sigma^2$ we have divided by $|V_{org}|$. 
Here we can see that the $\chi^2$ is lowest when the strength of the artificial source is about $40\%$ of the Sun signal. % where the recovery of the artificial source best. 
When the source strength is small, the $\chi^2$ is dominated by the noise and the $\chi^2$ is high. On the other hand, when the strength of the artificial sources is high compared to the Sun contamination as described before, the recovery gets worse. 

\section{Discussion}

While developing AlgoSCR we explored multiple techniques and came across various issues. Here we discuss some of the points that are important in the context of optimizing AlgoSCR.  

In Sec.~\ref{sec:leftover}, we address the issue of removing the residual Sun signal after subtraction of the largest eigenvalue.  Here we introduce the concept of the multiplication factor $g$. This procedure raises the question of what will happen if instead of filtering out just the largest eigenvalue, we filter out a smooth component from the two largest eigenvalues. %We have also tried to address that issue during our analysis. 
We find that removing the two largest components after smoothing, then the signal from the second largest component, which includes some radio sources, also gets removed, i.e. we will be removing the components from other radio sources and hence the method will not work. 

%In Section 4.2
In Sec.~\ref{sec:scaling}, while choosing the gain values, $g$, we calculate the gains at intervals of $15$ min and then interpolate. 
%Our analysis shows that if we change the gain points, i.e. if instead of setting the gains every 15 min, we calculate the gains every 10 min or in 30 min, then it will not have much effect on the data as the variation is pretty smooth throughout the day.
We find that the results are fairly insensitive to the choice of time interval (say, 10 min or 30 min), as is expected because the gain varies smoothly throughout the day.
However, if we choose a long time interval (several hours) for setting the gains, then we expect the results to worsen as the gain may change significantly in that time. However, we have not simulated these cases. 

%Here, we should also note that even if we over-subtract or under-subtract the Sun signal while fully removing the largest eigenvalue, it does not matter as the gain factor applied in the later stage actually compensate for the over- or under-subtracted value. 

Another important fact that came up during our analysis is that the Sun removal algorithm works better with more baselines, i.e. if we use all $16$ dishes from Tianlai instead of, say, $10$ dishes, then the effectiveness of the algorithm increases. Analysis with fewer dishes increases the power leakage to other  eigenvalues. 
The exact reason behind this is not known, but it may happen as more baselines reduce the effect of the noise in the eigen-decomposition. 

In addition, if we increase the integration time from 1 second to a larger value, the results get worse, which may be due to the fact that the Sun is not a point source. This is different from co-adding the signal from multiple days, which is eventually what the Tianlai array is designed to do. However, we have not tested the algorithm on co-added signals yet. 

Our analysis shows that AlgoSCR removed most of the solar contamination during the day.
%The sun removed visibility is good enough for making map of the sky during the daytime. In the future, we want to insert an artificial HI signal with a known power spectrum, apply AlgoSCR to remove the Sun and other techniques to remove the other foregrounds, and see whether we recover the correct power spectrum for the HI.
However, it is just a first step. We have not yet tested its effect on map-making and power-spectrum estimation. A critical next step is to make sure that AlgoSCR does not affect the statistics of the maps. This can be checked by comparing the HI power spectra and other statistical quantities from the maps produced using only nighttime data and the maps produced using the full day data after solar contamination removal. Such an analysis requires foreground subtraction and mapmaking and is outside the scope of the present work.

\section{Conclusion}

%Several ongoing and upcoming experiments are planning to map the HI sky to map in 3D. They will provide invaluable information on both cosmology and astrophysics. However, in various radio interferometric arrays, daytime data are generally contaminated by the solar signal making it unusable  for any astronomical analysis, even though it makes up to half of the daily available data. %This is the case with Tianlai data, in which there is strong solar contamination during the day. 

%Per the radiometer equation, the rms signal from the receiver output is inversely proportional to the square root of the integration time $\tau$, so to detect the 21 cm fluctuation (in the order of a few mK), a receiver with a system temperature of 100K and bandwidth $\delta\nu=250$ kHz would requires over $10000$ hours of integration time. 
%underlying

In this paper, we present a way to separate out the solar contamination from the daytime data observed by an interferometric radio array using eigen-decomposition techniques. The technique is primarily based on the assumption that if the Sun signal is the dominant signal in the sky, along with other weaker sources, and if the signals from the different sources are not correlated, then in the eigen-decomposition of the visibility matrix, the largest eigenvalue is from the strongest source, i.e. the Sun. The eigenvector corresponding to the largest eigenvalue points in the direction of that source in the eigenspace. The technique should filter out this largest eigenvalue while retaining the signals from other sources in the sky. 

However, antenna gain fluctuations, noise, sidelobe gain patterns, ground reflection, thermal effects on the instruments and cables, and cross-talk between antennas introduce mixing between the largest eigenvalue and other smaller eigenvalues. For these reasons singling out and removing the Sun signal is not straightforward, and there is some residual contamination from the Sun. Therefore, we apply some novel techniques to remove the leftover Sun signal.  

We show that our algorithm is able to remove the solar contamination without removing other, weaker sources in the sky. We have also presented the application of our algorithm to simulated signals, where the background radio sky is known. We show that AlogSCR can recover the background signal pretty well, after removing the solar contamination. %%% to prove the efficacy of the algorithm. 

To the best of our knowledge, this is the first published method for removing solar contamination from radio interferometer data. AlgoSCR can contribute to other ongoing and upcoming radio interferometers for solar contamination removal.

%However, it is just a first step; we have not yet tested its effect on map-making and power-spectrum estimation. A critical next step that we plan is to make sure that AlgoSCR does not affect the statistics of the maps. 

\section{Acknowledgement}

The authors also wish to thank Richard Shaw, Jeff Peterson and John Marriner for several fruitful discussions about the project during the Tianlai meeting in Guizhou. We wish to thank Kevin Gayley for allowing us to use the beam pattern from his CST simulation. We thank Juyong Zhang and his team for letting us use the beam mapping data from their drone survey. 

Work at UW-Madison and Fermilab is partially supported by NSF Award AST-1616554.

This research was performed using the compute resources and assistance of the UW-Madison Center For High Throughput Computing (CHTC) in the Department of Computer Sciences. The CHTC is supported by UW-Madison, the Advanced Computing Initiative, the Wisconsin Alumni Research Foundation, the Wisconsin Institutes for Discovery, and the National Science Foundation, and is an active member of the Open Science Grid, which is supported by the National Science Foundation and the U.S. Department of Energy's Office of Science.  

The Tianlai array is operated with the support of NAOC Astronomical Technology Center. The work at NAOC is supported by the Ministry of Science and Technology of China under grants 2018YFE0120800, 2016YFE0100300 and 2012AA121701, the National Natural Science Foundation of China under grants 11633004, 11473044, 11761141012, 11653003, 11773031, Chinese Academy of Science grants QYZDJ-SSW-SLH017, XDA15020200, ZDKYYQ20200008.

This document prepared by the Tianlai Collaboration includes personnel and uses resources of the Fermi National Accelerator Laboratory (Fermilab), a U.S. Department of Energy, Office of Science, HEP User Facility. Fermilab is managed by Fermi Research Alliance, LLC (FRA), acting under Contract No. DE-AC02-07CH11359.

\appendix

\section{Summary of AlgoSCR}
\label{App:1} 

Here, we review the step-by-step procedure for Sun removal using the algorithm described above. 

\begin{itemize}
\item For this procedure to work, first we separate the visibility $\mathbf{V}$ into the horizontal and vertical polarizations, $\mathbf{V}^{(H)}$ and $\mathbf{V}^{(V)}$, respectively. 
%We will drop the $(t,\nu)$ from now on and it's understood that each visibility matrix contains the visibilities at a particular time $t$ and frequency $\nu$. 
If we don't separate the polarizations, the noise and crosstalk in the same dish will give an additional large eigenvalue. The dimension of $\mathbf{V}$ is $32\times 32$, since we have 16 dual-polarization feeds. The dimension of $\mathbf{V}^{(H)}$ and $\mathbf{V}^{(V)}$ will be $16\times 16$.

\item Remove the night-time mean from the visibility matrix $\mathbf{V}^{(X)}$: $\mathbf{V}^{(X)}=\mathbf{V}^{(X)}-\langle\mathbf{V}^{(X)}\rangle_\mathrm{night}$. Here the average is over the time direction for different frequency channels. This will remove the cross-talk between the antennas.% and 
%$\mathbf{V}^{(V)}=\mathbf{V}^{(V)}-\langle\mathbf{V}^{(V)}\rangle_\mathrm{night}$. 
%This DC offset is discussed in Section~\ref{sec:DCOffset}. For the rest of the analysis, $\mathbf{V}^{(X)}$ is to be understood as the night-time mean average removed visibility.

\item Replace the auto-correlations by Eq.~\ref{Eq.autocorrelationRenorm}. In practice, if the denominator, $\mathbf{V}^{(X)}_{(i,j)}$, is zero, we replace the term inside the sum by a small number, such as $0.0001$.

\item Perform an eigen-decomposition of $\mathbf{V}^{(X)}$:

\begin{eqnarray}
\mathbf{V}^{(X)} = \mathbfcal{E}^{(X)} \mathbf{\Lambda}^{(X)} (\mathbfcal{E}^{(X)})^{-1}. %\\
%\mathbf{V}^{(V)} = \mathbfcal{E}^{(V)} \mathbf{\Lambda^{(V)}} (\mathbfcal{E}^{(V)})^{-1}
\end{eqnarray}

\item For each second of integration time, let the largest (normalized) eigenvector corresponding to the largest eigenvalue, $\lambda^{(X)}_{S\,(t,\nu)}$, be $\mathcal{E}^{(X)}_{S\,(t,\nu)}$. Now $\mathcal{E}^{(X)}_{S\,(t,\nu)}$ is a vector containing $n = 16$ complex numbers.  For fitting the smooth line through these vectors we calculate the tangents, $T^{(X)}_{S\,(t,\nu)}$ as: %Without loss of generality, we only show the procedure for the $x$-polarization, since the procedure for the $y$-polarization is similar.

\begin{equation}\label{tangents}
T^{(X)}_{S\,(t,\nu)}(i) = \frac{\|\mathcal{E}^{(X)}_{S\,(t,\nu)}(i)\|}{\sqrt{\displaystyle{\sum_{j=i+1}^{n}{\left(\|\mathcal{E}^{(X)}_{S\,(t,\nu)}(j)\|\right)^2}}}} \hspace{.5cm} \forall i \in [1,n-1].
\end{equation}

\item 
For smoothing the tangents, $T^{(X)}_{S\,(t,\nu)}$, along the time direction, we apply a Butterworth low-pass filter to remove the high frequency signal. %The parameters for the Butterworth filter is chosen empirically. 
For our dataset, the filter order is 2 and the $-3$ dB cut-off frequency is $0.01$ Hz. A $0$ phase filtering is done by \texttt{scipy}'s \texttt{filtfilt} function. Let the filtered (smoothed) tangents be $\tilde{T}^{(X)}_{S\,(t,\nu)}$.

\item Convert the eigenvectors back to Cartesian coordinates.
For each second of integration time,
\begin{eqnarray}
\| \tilde{\mathcal{E}}^{(X)}_{S\,(t,\nu)} (i) \| = \sin\left(\tan^{-1}(\tilde{T}^{(X)}_{S\,(t,\nu)}(i))\right)\times  \nonumber\\
\prod_{j=1}^{i} \cos\left(\tan^{-1}(\tilde{T}^{(X)}_{S\,(t,\nu)}(j))\right), \qquad
    \forall i\in [1,n-1]
\end{eqnarray}
 and 
\begin{equation}
    \| \tilde{\mathcal{E}}^{(X)}_{S\,(t,\nu)} (n)\| = \prod_{j=1}^{n} \cos\left(\tan^{-1}(\tilde{T}^{(X)}_{S\,(t,\nu)}(j))\right)\,.
\end{equation}
We then calculate the real and the imaginary parts of the  of the eigenvectors
\begin{equation}
    \tilde{\mathcal{E}}^{(X)}_{S\,(t,\nu)}(i) = \| \tilde{\mathcal{E}}^{(X)}_{S\,(t,\nu)}(i) \|\Big[\cos\left(\theta^{(X)}_{S\,(t,\nu)}(i)\right) + i \sin\left(\theta^{(X)}_{S\,(t,\nu)}(i)\right)\Big]
\end{equation}
where $\theta^{(X)}_{S\,(t,\nu)}(i)=\tan^{-1}\left(\Im(\mathcal{E}^{(X)}_{S\,(t,\nu)})/\Re(\mathcal{E}^{(H)}_{S\,(t,\nu)})\right)$.

\item The contribution to the visibility from the Sun is given by
\begin{equation}\label{vis_Sun}
    \mathbf{V}^{(X)}_{S\,(t,\nu)} = \lambda^{(X)}_{S\,(t,\nu)}\left(\tilde{\mathcal{E}}^{(X)}_{S\,(t,\nu)} \otimes \tilde{\mathcal{E}}^{(X)}_{S\,(t,\nu)} \right)
\end{equation}
where $\otimes$ denotes the outer product between eigenvector $\mathcal{E}^{(X)}_S$ and itself.

\item After removing the Sun contribution, the sky contribution to the visibility is 
\begin{eqnarray}
    \mathbf{V}^{(X)}_{Sky} &=& \mathbf{V}^{(X)} - \mathbf{V}^{(X)}_S \nonumber\\
    &=& \mathbfcal{E}^{(X)} \mathbf{\Lambda}^{(X)} (\mathbfcal{E}^{(X)})^{-1} - \mathbf{V}^{(X)}_S.
\end{eqnarray}
However, the above steps still leave some Sun signal contamination, as shown in Fig.~\ref{fig:zero_eigseparation}. To better remove this leftover Sun contamination, we multiply the Sun signal in Equation \ref{vis_Sun} 
by a complex factor of $g=A e^{i\phi}$:
\begin{equation}
\mathbf{V}^{(X)}_{Sky} = \mathbf{V}^{(X)} - A e^{i\phi}\mathbf{V}^{(X)}_S.
\end{equation}

\item To find $A$ and $\phi$, we divide the 10 hours of daytime data into 36 intervals (1000 s each) and minimize the $\chi^2$ for each interval. Here we define the  $\chi^2$ as
\begin{eqnarray}
    \chi^2 = \sum_{t,\nu} \left(\Re\left[\mathbf{V}^{(H)} - A e^{i\phi}\mathbf{V}^{(H)}_{S\,(t,\nu)}\right]\right)^2 + \nonumber\\ \sum_{t,\nu} \left(\Im\left[\mathbf{V}^{(H)} - A e^{i\phi}\mathbf{V}^{(H)}_{S\,(t,\nu)}\right]\right)^2
\end{eqnarray}
The sum is done over all seconds in the chosen interval and frequency $700.625~\mathrm{MHz}$ to $794.375~\mathrm{MHz}$. We don't sum the frequency channels before $700~\mathrm{MHz}$ and after $800~\mathrm{MHz}$, because we don't want to include the edge of the band-pass filter.
At the end of this process, we have 36 $[A,\phi]$ pairs. % In doing so, we minimize the term for each time interval. 

\item We have 36 $[A,\phi]$ pairs corresponding to thirty-six 1000 sec intervals in 10 hours of daytime data. We use a cubic spline to interpolate a $[A,\phi]$ pair for each second in 10 hours of daytime data.

\item Subtract the corrected Sun signal:
\begin{equation}
\mathbf{V}^{(H)}_{\text{Sky}\,(t,\nu)} = \mathbf{V}^{(H)} - A_{\text{int}\,(t,\nu)} e^{i\phi_{\text{int}\,(t,\nu)}}\mathbf{V}^{(H)}_{\text{S}\,(t,\nu)}.
\end{equation}
This gives us the final solar contamination removed result, and the results are shown in Fig.~\ref{fig:chisquare_gain}.

\end{itemize}

\section{Calculating the artificial source visibilities}
\label{App:2} 
%\begin{figure}
%    \centering
%    \includegraphics[width=0.8\textwidth]{figs/CHTC.pdf}
%    \caption{Example of the workflow run on UW-Madison Center For High Throughput Computing (CHTC)
%    }
%    \label{fig:bl40_zeroeig}
%\end{figure}

The visibilities of the artificial sources come from three made up sources near the NCP. %with declination are generated for each frequency similar to that of the Tianlai Dish Array. 
The three simulated sources are randomly chosen to be at (RA, DEC) = (75.75,81.25), (79.5, 80.5) and (245.0,79,75) with constant brightness temperatures across all observed frequencies.
%, respectively. 
We calculated the visibilty for each frequency in Tianlai's 512 frequency bins (equally spaced between 685 MHz and 810 MHz).

%amplitude 546, 170, 128 Jansky
% The simulated beam profile is assumed to have a Gaussian profile for declination greater than 75° and an flat sidelobe gain $10^-{2.8}$ for declination less than 75.

% The Tianlai Dish Array antenna positions were measured in a coordinate system of East, North, and Up $(E,N,U)$ reference frame with respect to the local horizon. Since the Tianlai Arrays are located at a latitude of $\mathcal{L}=$. Baselines in equatorial $(X,Y,Z)$ coordinates to a reference antenna (not in wavelength unit) can be found by a simple rotation matrix, with $\mathcal{L}$ is the latitude
% \begin{equation}
% \begin{pmatrix}
%     X \\
%     Y \\
%     Z \\
% \end{pmatrix}
% =
% \begin{pmatrix}
% 0 & -\sin \mathcal{L} & \cos \mathcal{L} \\
% 1 & 0 & 0 \\
% 0 & \cos \mathcal{L} & \sin \mathcal{L} \\
% \end{pmatrix}
% \begin{pmatrix}
%     E \\
%     N \\
%     U \\
% \end{pmatrix}
% \end{equation}

Each astronomical source exhibits a linearly varying phase with time and frequency, since the visibility is proportional to $e^{-i\varphi}$, where $\varphi$ is the fringe phase and is defined as
\begin{equation}
    \varphi=2\pi\nu\tau_g(\nu,t)=\frac{2\pi\nu\mathbf{b}\cdot\mathbf{s}}{c}\,.
\end{equation}

\begin{figure*}
    \centering
    \includegraphics[width=0.45\textwidth,trim = 0 1 1 1, clip]{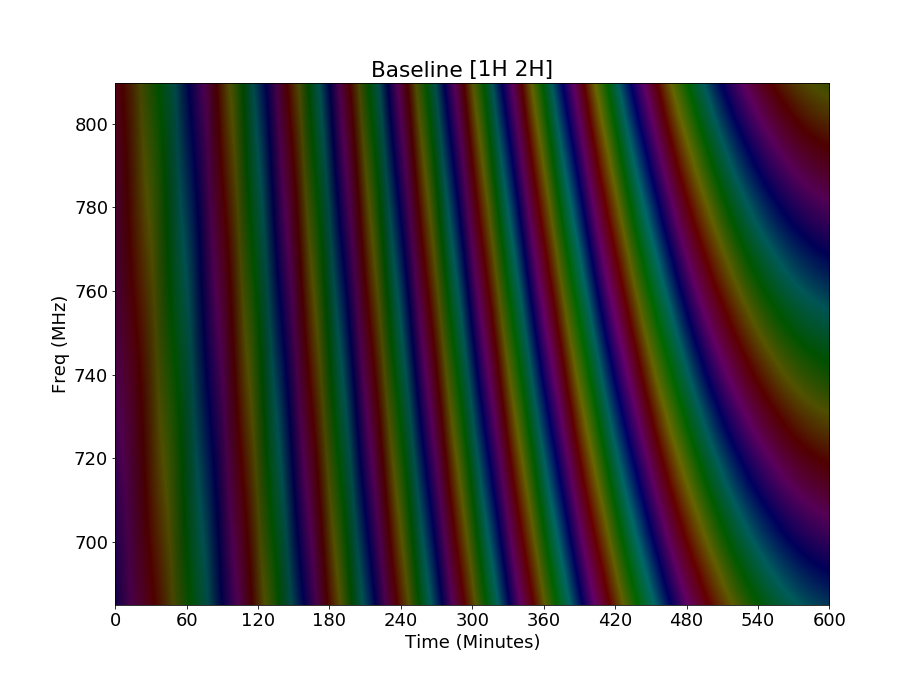}
    \includegraphics[width=0.45\textwidth,trim = 0 1 1 1, clip]{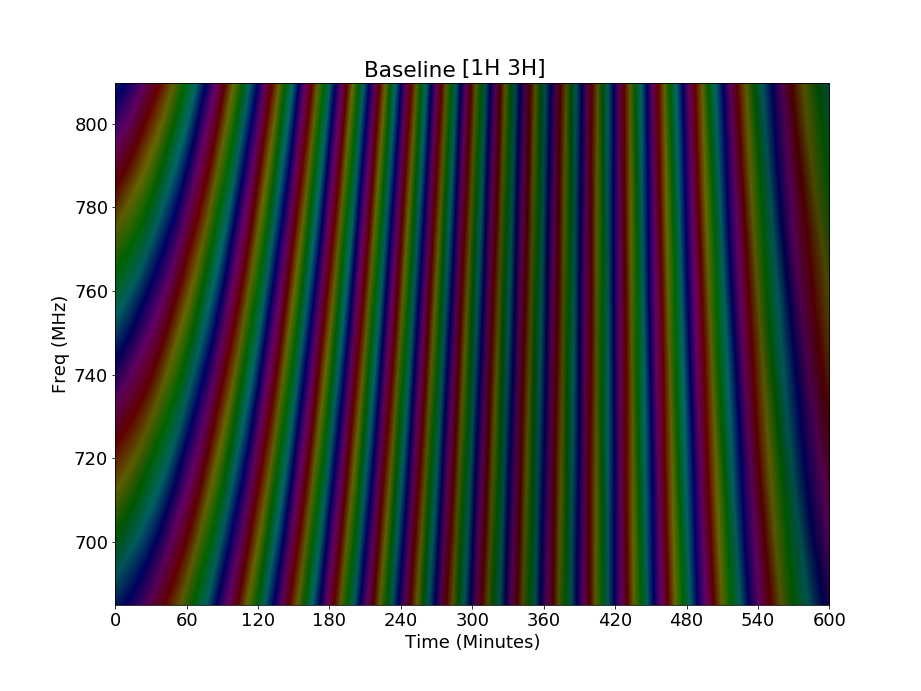}
    \includegraphics[width=0.45\textwidth,trim = 0 1 1 1, clip]{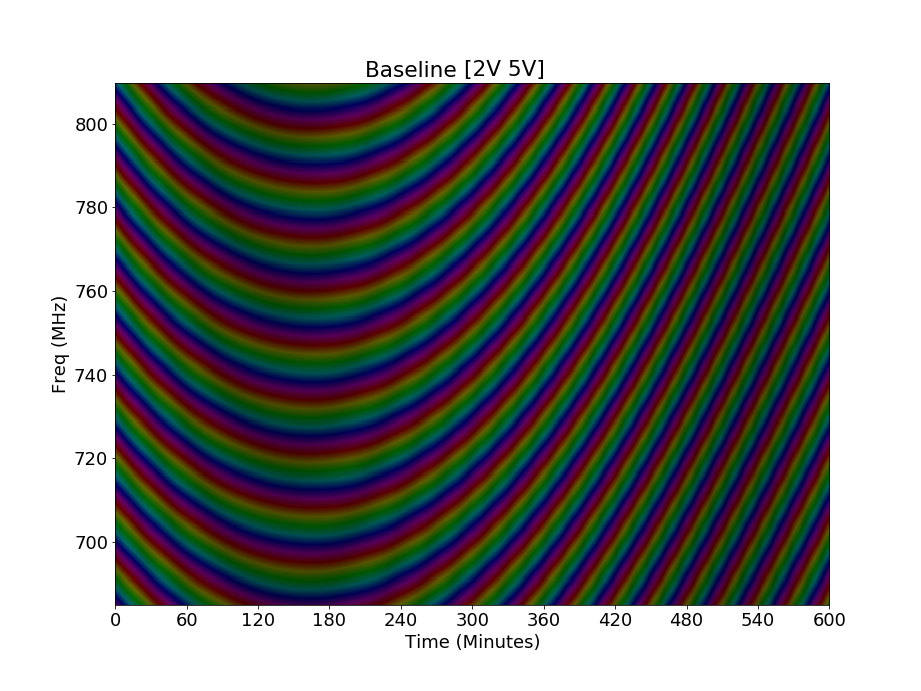}
    \includegraphics[width=0.45\textwidth,trim = 0 1 1 1, clip]{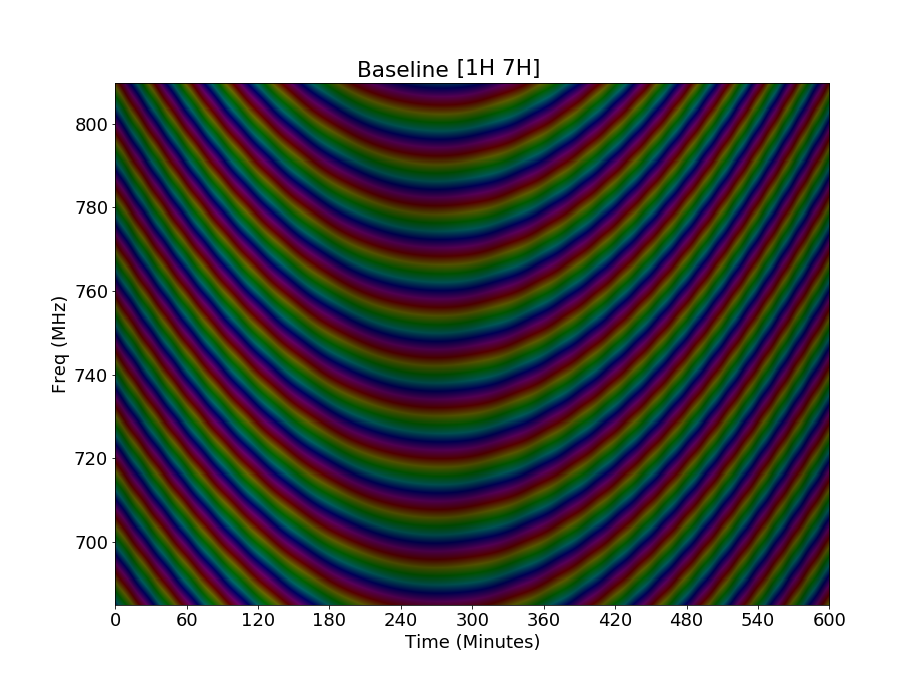}
    \caption{
    Combined visibilities of three artificial sources for different baselines, shown here for 10 hours. As expected, longer baselines give higher fringe rates, and for a given baseline, we see a faster fringe rate at lower frequency.
    }
    \label{fig:simulated_vis}
\end{figure*}

\noindent $\tau_g(\nu,t)$ is the frequency independent geometric delay and is equal to
\begin{eqnarray}
    \tau_g(\nu,t)&\equiv& \frac{\mathbf{b}\cdot\mathbf{s}}{c} 
    =\frac{b_x}{c}\cos{\delta}\cos{H(t)}\nonumber \\
    &&-\frac{b_y}{c}\cos{\delta}\sin{H(t)}+\frac{b_z}{c}\sin{\delta},
\end{eqnarray}

\noindent where $\delta$ is the source declination, $H(t)$ is the source hour angle as a function of sidereal time, $\mathbf{b}=(b_x,b_y,b_z)$ are the baseline components with units of length in the radial, eastern and northern polar directions and $\mathbf{s}$ is the source vector. $c$ is the speed of light. We can also calculate the fringe rates as follows:
\begin{equation}
    \frac{\partial{\varphi}}{\partial t}=\frac{2\pi\nu}{c}[-b_x\sin{H(t)}+b_y\cos{H(t)}]\cos{\delta},
\end{equation}
For each second of integration time and each frequency, the visiblity for dish $i$ and $j$ is calculated as follows
\begin{equation}
    \mathbf{V}_{(i,j)}=\sum_{k=1}^n \mathcal{F}_k A(\mathbf{s}) e^{-i\varphi}
\end{equation}
where $\mathcal{F}_k$ is the flux of source $k$. In our simulations, we used three sources with $\mathcal{F}=(546,170,128)$ Jansky. $A(\mathbf{s})$ is the gain of the antenna in the direction of the source vector $\mathbf{s}$. For simplicity, $A(\mathbf{s})$, the simulated main beam gain, is taken as a Gaussian distribution with a standard deviation of $3^\circ$ (FWHM = $7^\circ$), and we assume that all three artificial sources fall within the main beam.  Therefore we did not model the beam sidelobe gain. $A(s)$ is also assumed to be independent of frequency. In the real Tianlai Dish beam pattern, the mainbeam (excluding the sidelobe) FWHM is about $5^{\circ}$ (\cite{wu2020tianlai}). The simulated baselines are identical to the real Tianlai dish array, and the procedure for calculating the visibility is repeated for every baseline. The waterfall plots of the simulated visibilities for a few typical baselines are shown in Fig.~\ref{fig:simulated_vis}. As expected, longer baselines give higher fringe rates, and for a given baseline, we see a faster fringe rate at lower frequency.

%, since the Tianlai Dish Array beamwidth is about $5^\circ-6^\circ$ (\cite{wu2020tianlai}).

\begin{figure*}
    \centering
    \includegraphics[width=0.45\textwidth,trim = 0 1 1 1, clip]{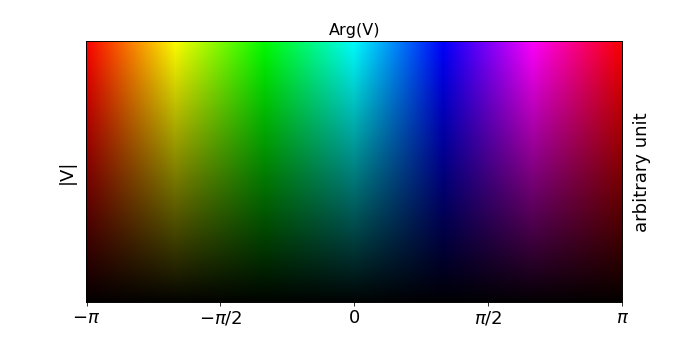}
    \caption{
    The color palette used to represent complex visibilities in this paper is shown in this plot. The phase of the complex visibility is represented by the color and the amplitude of the visibilities are represented by the hue of the color. 
    }
    \label{fig:palette}
\end{figure*}

\bibliographystyle{plainnat}
\bibliography{reference}

\newpage
\end{document}